\documentclass[a4paper]{article}

\usepackage[numbers,sort&compress]{natbib}
\usepackage{graphicx}
\usepackage{xcolor}
\usepackage{amsmath,amssymb}
\usepackage{caption}
\usepackage{url}

\begin{document}

\begin{center}
    {\LARGE \bfseries Impact of boiling liquid droplets:\\ 
    Vapor entrapment suppression \par}
    \vspace{1em}
    Bernardo Palacios-Mu\~{n}iz, Edgar Ortega-Roano, Yee Li (Ellis) Fan,\\
    Nayoung Kim, and Devaraj van der Meer, \par
    \vspace{0.5em}
    {\small\it
      Physics of Fluids Group, Max Planck Center Twente for Complex Fluid Dynamics,\\ University of Twente, 7500 AE Enschede, The Netherlands.
    }\par
    \vspace{1em}
    \textbf{Abstract} \par
    \vspace{1em}
    \begin{minipage}{0.85\textwidth}   
      \noindent
      {\small There hardly is a fluid mechanics phenomenon attracting more attention than the impact of a droplet, due to its undeniable beauty, many applications and the numerous challenges it poses. One of the crucial factors turns out to be the cushioning effect of the gas surrounding the droplet. This fact, together with the observation that almost all of the relevant literature was done in air, triggers the question what would happen when the liquid was a boiling liquid, i.e., a liquid in thermal equilibrium with its own vapor, as is the case during transport of cryogenic liquids such as liquid hydrogen. To investigate precisely this question, we experimentally generate droplets in thermodynamical equilibrium with their own vapor, even before impact, such that minute energy exchanges of the droplet with its surroundings can trigger phase change. Using a frustrated total internal reflection (TIR) setup, we make the exciting observation that depending on the impact speed and vapor conditions, the entrapment of vapor can be completely suppressed under boiling liquid conditions. We create a simplified model based on scaling arguments and perform numerical simulations considering both the compressible and condensable properties of the vapor layer that are in very good agreement with our experimental findings. Our results can be of great consequence to the pressures exerted during droplet impact and on an industrial scale may help better understand the loads experienced during sloshing wave impact inside cryogenic liquid containers.}
    \end{minipage}
    \noindent
  \end{center}

\noindent Droplet impact has long remained a subject of interest to fluid mechanics for its beauty, applications and numerous challenges \cite{Yarin2006, Josserand2016DropSurface}. In fact, when it comes to studying impacting droplets, several factors can affect the sequence of events which will unfold, such as the substrate's roughness \cite{Kolinski2014DropsSurfaces,Garcia-Geijo2021SpreadingSubstrates}, or  whether the liquid wets it or not \cite{Quetzeri-Santiago2019} and even the ambient pressure, which can suppress the characteristic splash of high velocity droplet impacts \cite{Xu2005DropSurface}. After the discovered relevance of the ambient pressure, of course related to the gas surrounding the droplet, recent attention has been brought to the previously neglected interaction of the droplet with the gas layer between the liquid and solid which has led to many relevant findings. Roughly speaking, when a falling droplet 
gets close enough to a surface, pressure builds up in the gas layer below the droplet's bottom surface and the gas underneath is squeezed out radially. This leads to the formation of a dimple shape along the droplet's lower surface  \cite{Lohse2022FundamentalPrinting} which will have relevant consequences later in the impact. One of these is that the very early interaction with the gas layer can later lead to splashing \cite{Mandre2012TheSurface,Riboux2014ExperimentsSplashing}, another is that the dimple shape can become an entrapped gas bubble after touchdown \cite{Bouwhuis2012MaximalImpact} or it can 
lead to jetting, either upwards \cite{Lee2011SizeBursting} or downwards \cite{Lee2020DownwardDrop}, which in turn may also change the dynamic impact force on the surface \cite{Zhang2022ImpactSurfaces}. With all this in mind, it can be easily concluded that the gas layer interaction still holds great importance in the research of impacting droplets and its many applications. 

In this work we replace gas by vapor, by focusing on the impact of boiling liquid droplets, that is, droplets that are at their boiling point, in thermal equilibrium with the vapor that surrounds them. More specifically, we want to focus on the entrapment of vapor inside the dimple formed below the droplet due to pressure accumulation. A related subject is that of droplets impacting in (or close to) the well-known Leidenfrost regime \cite{Quere2013LeidenfrostDynamics}, where due to evaporation a droplet can levitate over a heated solid upon impact. These have been performed in atmospheric 
\cite{Tran2012DropSurfaces,lee2018transient,gordillo2022initial,Chantelot2023DropDominance,garcia2024skating,sprittles2024gas}
and reduced pressure conditions \cite{VanLimbeek2018a}. Unlike this previous work, we do not use large temperature differences to trigger rapid phase change, since the liquid will be in thermodynamical equilibrium with its own vapor, even before impact. When in such a state, minute energy exchanges of the droplet with its surroundings can trigger phase change \cite{ezeta2025large}. Moreover, unlike the Leidenfrost impact work where phase change is limited to the evaporation of the liquid due to high substrate temperatures, in this boiling liquid state the condensation of vapor due to pressure locally rising above the vapor pressure of the liquid is also possible. We will show that the latter may lead to dramatic effects.
 
This study is curiosity driven, but with a specific application in mind: During the maritime transport of cryogenic liquids, such as liquid natural gas (LNG) or liquid hydrogen (LH2), a liquefied gas is loaded onto a specialized carrier and hauled for long periods of time. Naturally, such a liquid is what we have defined above to be a boiling liquid: a liquid at its boiling point in thermal equilibrium with its own vapor. One of the main concerns during transport is to insulate the main container, in order to keep a low boil-off rate of the liquid to
avoid losses of product. In this engineering challenge, the trade off for good insulation is loss of structural integrity,  turning loads due to liquid impact inside of the container into a major concern \cite{dias2018slamming}. While LNG carrier design has now long been known to work in a safe manner \cite{Vanem2008AnalysingOperations}, the loads within it are not well understood, which is especially true for possible effects of phase change. This work aims to help understand such loads, by reducing the problem to the more fundamental one of droplet impact in equilibrium conditions that reflect those within cryogenic containers.
Further research can help the transition to more sustainable fuels, such as LH2, which will pose further, more complex challenges. 

\subsection*{Experimental setup}

\begin{figure}
\centering
\includegraphics[width=0.85\linewidth]{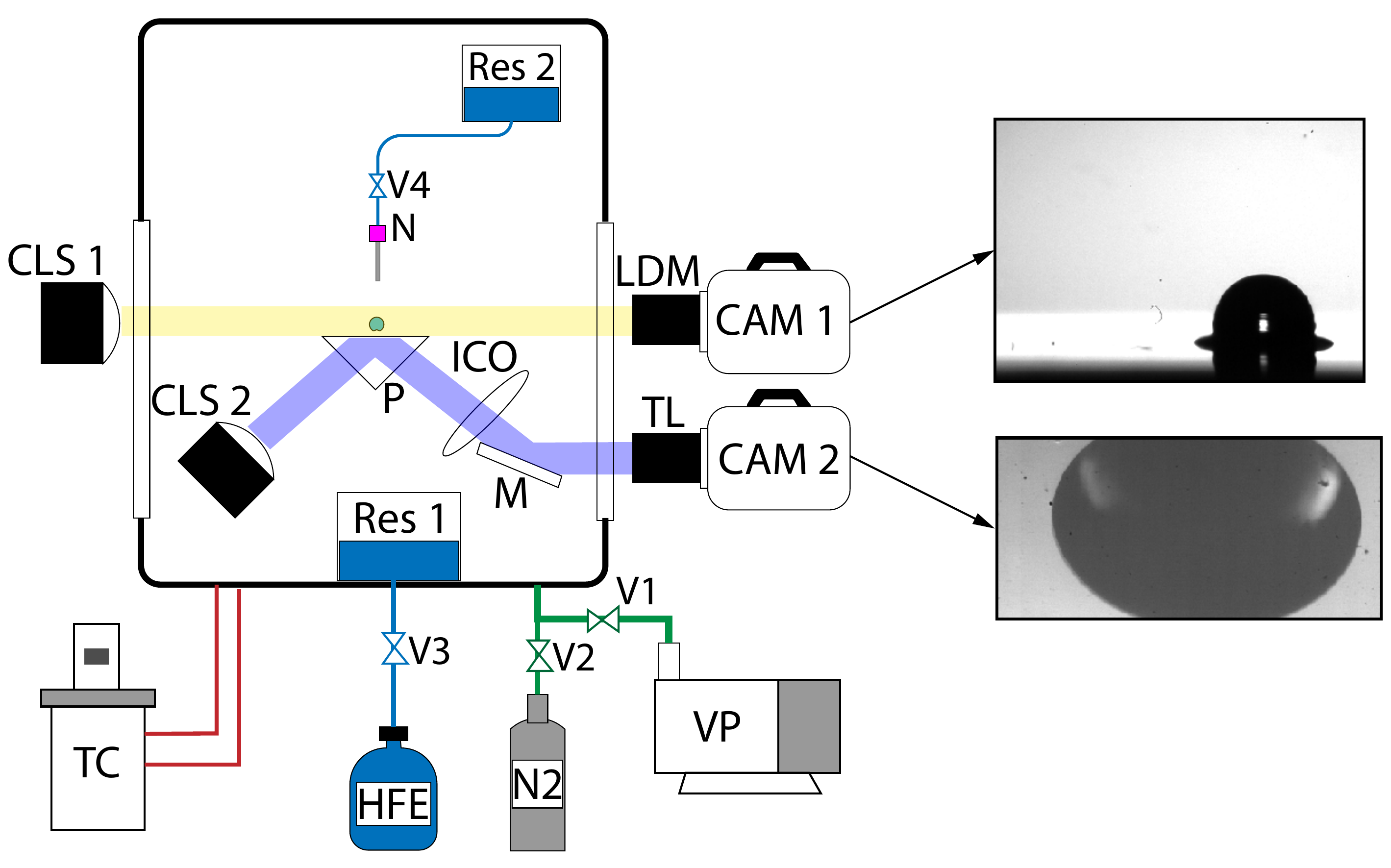}
\caption{\label{fig:setup_sketch} Sketch of the experimental setup with as main elements the closed, temperature-controlled chamber containing HFE-7000 liquid (Res1) in equilibrium with its vapor, the droplet generated from the second reservoir (Res2), and the prism (P) used for frustrated total internal reflection imaging (TIR), the vacuum pump (VP), and the two high-speed cameras (CAM1,2) used for imaging.} 
\end{figure}

In order to generate droplets in equilibrium with their own vapor and record their impact onto a substrate, we have developed a highly versatile setup, using HFE-7000 as a working fluid, which has an appreciable vapor pressure ($P_{V,0} = 0.57$ bar) at room temperature. 
Here we provide a general description of the setup (see Fig.~\ref{fig:setup_sketch}),
while a more detailed description can be found in the Supporting Information (SI).

First, a saturated liquid-vapor environment at temperature $T_0 \approx 22$ $^\circ$C is created by allowing HFE-7000 to evaporate inside a hermetically sealed, vacated, and temperature-controlled chamber. 
Enough liquid is allowed into a reservoir (Res1) to ensure the chamber remains saturated with vapor during the entire experiment. 
Additional HFE-7000 liquid is allowed into a smaller reservoir (Res2), located higher in the setup, connected to a needle (N) with its tip slightly below the liquid level, allowing droplets to fall down and impact a sapphire prism (P). 
The height between needle and substrate can be varied to obtain an impact velocity $U_0$ between $0.2$ m/s and $1.5$ m/s. The substrate temperature $T_s$ is controlled between $T_0$ and $40$ $^\circ$C. 

Side views of the impact are recorded at 16,000 fps and are used to measure the impact speed $U_0$ and  initial radius $R_0$ of the droplets. Bottom views are captured at frame rates between $96,000$ and $200,000$ fps by illuminating the top surface of the prism internally with a collimated monochromatic light source (CLS2) from below, such that it totally reflects. If a droplet comes within the evanescent wave distance from the surface, then light is partly, and upon contact fully transmitted into the droplet, which results in dark regions on the otherwise bright surface. This technique is known as frustrated total internal reflection (TIR) and has been extensively used in the past 
\cite{Shirota2016, Kolinski2014Lift-OffSurface,Chantelot2023DropDominance}.

\subsection*{Results and discussion}
We have performed an extensive experimental campaign, divided in two parts. For the first part we have performed experiments where boiling droplets, falling at different velocities, impact a slightly heated surface. The latter is done to avoid the formation of a liquid film on top of the substrate that would interfere with the impact. For simplicity, we will refer to the experiments in this first part as (near-)isothermal experiments.

\subsection*{Near isothermal impact of boiling droplets}

\begin{figure}
\centering
\includegraphics[width=0.85\linewidth]{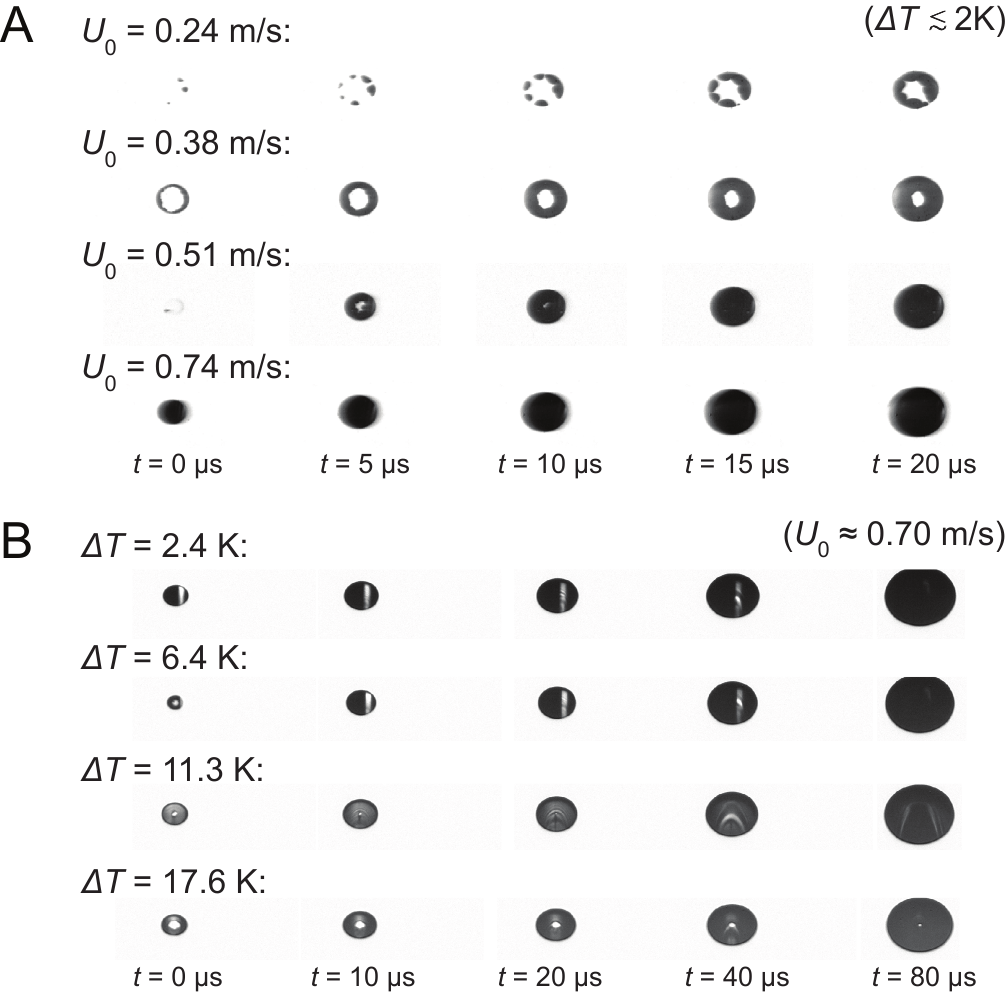}
\caption{\label{fig:isothermal_evo} (\textit{A}) TIR image sequence (after background subtraction) of a series of HFE-7000 droplets with radius $R_0 \approx 0.6$ mm in a vapor atmosphere at $T_{0} \approx 22$ $^\circ$C; the droplets impact a sapphire substrate heated to a temperature difference  $\Delta T \lesssim 2$ $^\circ$C with impact velocities $U_0 = 0.24$, $0.38$, $0.51$, $0.74$ m/s, Fig. S2 and movies S1 to S5 in the SI. At low impact velocity a vapor bubble is entrapped, but at high impact velocities this entrapment is suppressed.
(\textit{B}) TIR image sequence (after background subtraction) of a series of droplets with radius $R_0 \approx 0.6$ mm in a vapor atmosphere at $T_0 \approx 22$ $^\circ$C impacting with velocity $U_0 \approx 0.7$ m/s onto a sapphire substrate heated to an increasing temperature difference $\Delta T = 2.4$, $6.4$, $11.3$, $17.6$ K (with respect to $T_0$), Movies S6 to S10 in the SI. At this impact velocity, the entrapment of vapor is suppressed for small $\Delta T$, but entrapment can be recovered by heating the substrate to a higher temperature. 
In all cases, time is measured with respect to the first TIR frame in which the droplet becomes visible.}
\end{figure}

Fig.~\ref{fig:isothermal_evo}\textit{A} shows TIR snapshots from typical near-isothermal experiments for droplets impacting a substrate at different impact speeds. The frames are a TIR sequence where wet (dark) and dry (light) zones are well distinguished. For $U_0 = 0.38$ m/s, in the first frame, a dark ring appears over the background, corresponding to the surface of the droplet making contact with the substrate. The light part within the ring is the dimple which has formed below the droplet, indicating entrapped vapor. As the droplet further moves down, the ring expands radially, both inwards and outwards.

This matches the chain of events observed for experiments conducted when the surrounding fluid is a non-condensable gas, as described in several other works \cite{THORODDSEN2005TheSurface,Bouwhuis2012MaximalImpact,Kolinski2012SkatingSurface,Lee2020DownwardDrop,Zhang2022ImpactSurfaces,Kaviani2023CharacteristicMica}. So we stress that a boiling liquid also entraps the vapor surrounding it, at least at this small impact velocity. However, increasing the impact velocity can lead to a radical change in behavior, which we can observe for the larger impact velocities in Fig.~\ref{fig:isothermal_evo}\textit{A}. 

For the smallest impact velocities in this work ($U_0=0.24$ m/s in Fig.~\ref{fig:isothermal_evo}\textit{A}) the contact starts on $3$ separate locations from where the liquid starts spreading. The non-symmetrical first contact for this case can likely be attributed to 
capillary waves due to the pinch off from the needle, which at this short time distance have not yet dissipated. Despite the lost symmetry, soon after, more points initiate contact and after 20 $\mu$s vapor has been fully entrapped by the liquid (Fig. S2 in SI). As the fall height of the droplet increases, the capillary waves are no longer dominant and, from $U_0 = 0.38$ m/s onwards,
full entrapment of vapor within a wetted ring is recorded from the first frame, with both contact lines of the ring advancing. Increasing the impact velocity even more leads to significantly smaller amounts of vapor being entrapped and the inner contact line of the wet ring expands faster, leaving the entrapment visible for only a few frames at most, as seen for $U_0 = 0.51$ m/s in Fig.~\ref{fig:isothermal_evo}\textit{A}. A further increase of velocity can, strikingly, fully suppress the entrapment of vapor, as is observed for $U_0 = 0.74$ m/s in Fig.~\ref{fig:isothermal_evo}\textit{A}. To our knowledge, complete suppression of entrapment by means of an increasing impact velocity has not been reported before. 

\begin{figure}
\centering
\includegraphics[width=0.9\linewidth]{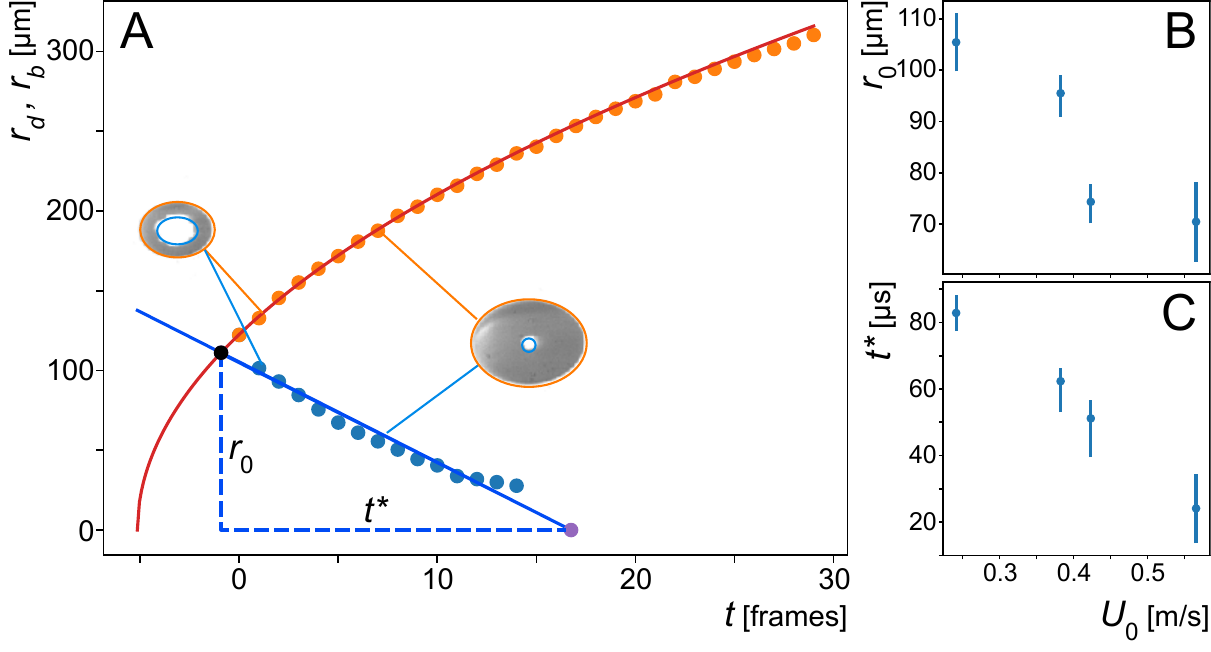}
\caption{\label{fig:contact_tracking} (\textit{A}) Azimuthally averaged inner ($r_b$, blue points) and outer ($r_d$, orange points) radius evolution for an experiment at $U_0 = 0.38$ m/s, with images available in the SI, Fig.~S1\textit{B}. The orange line is obtained from a linear fit to $r_d^2$ plotted against $t$ and the blue line is a linear fit for $r_b$. The intersection of the two curves provides the initial radius $r_0$ and also the initial impact time. The entrapment time $t^*$ is then measured from the inner radius fit as the time it takes for $r_b(t)$ to go from $r_0$ to $0$. 
(\textit{B}) Initial entrapment radius $r_0$ as a function of the impact velocity $U_0$ for a set of experiments with a small temperature difference of the impact surface with respect to the ambient temperature ($\Delta T = 1.77 \pm 0.49$ K). The points represent the mean of a minimum of three repetitions and the bars display the total error. (\textit{C}) entrapment time $t^*$ as a function of the impact velocity $U_0$, for the same set of experiments.}
\end{figure}

In order to better quantify the transition to the suppression of vapor entrapment we have measured the initial contact radius $r_0$ of the entrapped region, which is an indication of the size of the entrapped vapor bubble, and the entrapment time $t^*$, the time interval that the bubble is visible in the TIR images. To do this, we obtain the inner and outer radii 
$r_b$ and $r_d$ of the ring shape during the initial contact and track their time evolution, as exemplified in Fig.~\ref{fig:contact_tracking}\textit{A}. The outer radius $r_b$ is the spreading contact line of the wetting region, which is known to expand as $r_{b} \sim \sqrt{t}$ \cite{Riboux2014ExperimentsSplashing}. This behavior is fitted in Fig.~\ref{fig:contact_tracking}\textit{A} and shows excellent agreement to the data. The retracting contact line (inner radius) is approximated using a linear fit. 
We estimate the initial contact time and radius $r_0$ by determining the intersection point of the fitted curves. Additionally, we also find the zero crossing of the linear fit to the retracting inner contact line to obtain the vapor entrapment time $t^*$, which we measure with respect to the initial contact time (Fig.~\ref{fig:contact_tracking}\textit{A}).

The plots in Figs.~\ref{fig:contact_tracking}\textit{B} and~\ref{fig:contact_tracking}\textit{C} present the measured initial entrapment radius $r_0$ and entrapment time $t^*$, respectively, for the isothermal experiments. Each point in the plot is the mean value found for at least three repetitions and the vertical bars show the total error with respect to the mean. Both the initial entrapment radius and the entrapment time decrease monotonically with impact velocity, as was already clearly visible from Fig.~\ref{fig:isothermal_evo}\textit{A}. The initial contact radius is smaller than what has been previously reported for droplets surrounded by a non-condensable gas. It also presents a sudden decrease around $U_0 \approx 0.4$ m/s and is only measurable for impact velocities smaller than $0.6$ m/s. Impacts past this velocity result in no entrapment, such as for the largest velocity in Fig.~\ref{fig:isothermal_evo}\textit{A}. Most notably, at the threshold velocity around $0.6$ m/s, the entrapment time goes to a small asymptotic value that is consistent with zero, whereas the entrapment radius remains finite.

\subsection*{A simplified one-dimensional model}

In the next few paragraphs, through a simplified model, we find a quantitative explanation for the observed suppression of vapor bubble entrapment for increasing impact velocity. We begin considering an impacting droplet surrounded by a non-condensable gas. The pressure difference under such droplet is described by the lubrication flow of the gas as the droplet nears the surface \cite{Mandre2009PrecursorsSurface,Bouwhuis2012MaximalImpact}. As argued in that work, we can estimate the pressure build up below the droplet from the axisymmetric Stokes equation in the lubrication layer as
\begin{equation}
    \frac{\Delta P}{\ell} \sim \mu_g \frac{U_g}{h^2}\quad \Rightarrow\quad\Delta P \sim \mu_g \frac {R_0} {h^2} U_0 \,\,,
    \label{eq:pressurebuildup}
\end{equation}
where $\mu_g$ is the dynamic viscosity of the gas, $h$ is the thickness and $\ell \sim \sqrt{R_0h}$ the radial extent of the lubrication layer, where the latter is estimated geometrically  from the approach of a spherical droplet to a planar substrate, and the gas velocity $U_g$ is related to the impact velocity by continuity, $U_gh \sim U_0 \ell$. By equating this $\Delta P$ with the pressure needed to decelerate the liquid, namely $\rho_L(\partial\phi/\partial t) \sim \rho_LU_0^2\ell/h$ (with $\phi$ the flow potential), one obtains a scaling relation for the dimple height $h_d$ in terms of the Stokes number $\mathit{St} \equiv \rho_L U_0 R_0/\mu_g$
\begin{equation}
    \frac{h_d}{R_0} \sim \left(\frac{\rho_L U_0 R_0}{\mu_g}\right)^{-2/3} = \mathit{St}^{-2/3}\,\,,
    \label{eq:dimpleStokes}
\end{equation}
where $\rho_L$ is the density of the liquid. 

For our case, however, the surrounding phase is vapor and this means that an increase in pressure can trigger condensation: As soon as the pressure in the vapor layer rises, the vapor will try to condense to return to the equilibrium state. For simplicity, let's now consider a one-dimensional system composed of a solid substrate, an incoming liquid surface and an intermediate vapor layer. During pressure build up, vapor will try to condense and produce latent heat. Since simultaneously the temperature in the vapor rises, a thermal boundary layer will develop in the liquid, which will allow for transport of the latent heat into the liquid. Now, writing down the mass-energy balance at the liquid-vapor interface \cite{prosperetti2017vapor}, an approximation we also used in \cite{ezeta2025large,fan2025impact}
\begin{equation}\label{eq:massheatbalance}
    -L \frac{dm_V}{dt} = - S \, k_L \left.\frac{\partial T}{\partial r}\right|_\text{int} 
    \approx S \, k_L \frac{\Delta T_V}{\delta_{th}}\,\,,
\end{equation} 
where $L$ is the latent heat, $dm_V/dt$ the rate of change of vapor mass due to condensation, $S$ the surface area of the liquid-vapor interface, $k_L$ the thermal conductivity of the liquid, and $\partial T/\partial r|_\text{int}$ the temperature gradient at the interface. Note that we have neglected the amount of liquid produced during condensation due to the large liquid to vapor density ratio, and the sensible heat of the vapor in light of the much larger latent heat. Subsequently, we estimate the temperature gradient as the ratio of $\Delta T_V$, the difference between the temperature $T_V$ of the vapor and the equilibrium temperature $T_0$ of the liquid far away from the interface, and $\delta_{th}$, the thermal boundary layer thickness in the liquid phase. 

The condensation rate $-dm_V/dt$ of the vapor can be interpreted as a condensation speed in the vapor phase by dividing by vapor density $\rho_V$ and surface area $S$, i.e., $U_{cond} = - (dm_V/dt)/(\rho_V S)$, which after substituting in the previous equation gives us
\begin{equation}
    U_{cond}\approx \frac{k_L}{L \rho_V}\frac{\Delta T_V}{\delta_{th}}\,\,.
\end{equation} 
Assuming that the vapor bubble is in internal thermal equilibrium at all times (connected to neglecting the influence of sensible heat) the relation between vapor temperature and pressure is described by the Clausius-Clapeyron equation and the ideal gas law, for which one can write in linear approximation that
\begin{equation}\label{eq:CondSpeed}
    \frac{\Delta T_V}{T_0}\approx \beta \frac{\Delta P_V}{P_{V,0}}, \quad \text{with:}\quad \beta \equiv \frac{R_s T_0}{L}\,\,,
\end{equation}
where $R_s$ is the specific gas constant of the vapor phase and the equilibrium state variables of the vapor are denoted with subscript $0$: $P_{V,0}$, $\rho_{V,0}$, and $T_0 = T_{V,0}$. Finally, the boundary layer thickness $\delta_{th}$ can be approximated as $\delta_{th} \sim \sqrt{\pi \alpha_L \Delta t}$, with $\alpha_L$ the thermal diffusivity of the liquid and the time scale $\Delta t \approx \epsilon \frac{R_0}{U_0}$, where $\epsilon$ should be approximated from evaluating the time that the pressure rise in the lubrication layer starts being appreciable. 
This leads to $\epsilon \approx 0.02$, corresponding to $h\approx 0.02R_0$, which is justified in the SI and is both consistent with a similar estimate in \cite{gordillo2022initial} and with BI simulations (see next Section). Substituting everything into the expression for the condensation speed, and dividing by $U_0$ we obtain 
\begin{equation}
 \frac{U_{cond}}{U_0} \approx 
 \frac{\beta^2}{\sqrt{\pi\epsilon}}\frac{\rho_L}{\rho_{V,0}}\frac{c_{p,L}}{R_s}\sqrt{\frac{\alpha_L}{R_0U_0}}\frac{\Delta P_V}{P_{V,0}}\,\,,
\label{eq:Ucond}
\end{equation}
where we used $k_L = \alpha_L\rho_Lc_{p,L}$, with $\rho_L$ and $c_{p,L}$ the density and specific heat of the liquid. 

Now the main argument to connect the non-condensable gas derivation and the condensation speed derived above is the following: If at the moment that in a non-condensable gas situation the dimple would be formed, i.e., if \eqref{eq:pressurebuildup} and~\eqref{eq:dimpleStokes} would both hold, such that 
\begin{equation}
    \Delta P_V (\approx \Delta P_g) \sim \mu_V \frac {R_0}{h_d^2} U_0 \sim \frac{\mu_V U_0}{R_0}\mathit{St}^{4/3} \,\,,
    \label{eq:press}
\end{equation}
the condensation speed $U_{cond}$ is larger than the impact velocity $U_0$, then the vapor layer would condense faster than the impact speed and no vapor bubble will be entrapped. Conversely, if $U_{cond}<U_0$, we will see vapor bubble entrapment. Inserting \eqref{eq:press} into \eqref{eq:Ucond} leads to
\begin{equation}
\frac{U_{cond}}{U_0} \sim \frac{\beta^2}{\sqrt{\pi\epsilon}}\frac{\rho_L}{\rho_{V,0}}\frac{c_{p,L}}{R_s}\frac{\mathit{Eu}\,\mathit{St}^{1/3}}{\mathit{Pe}^{1/2}}\,\,,
\label{eq:final}
\end{equation}
where $\mathit{Pe} = R_0U_0/\alpha_L$ and $\mathit{Eu} = \rho_LU_0^2/P_{V,0}$ are the P{\'e}clet and the Euler numbers, respectively. Note that we have replaced $\mu_g$ with $\mu_V$ to clearly express that the dynamic viscosity of the vapor is intended.

Now, the threshold condition $U_{cond}/U_0 = 1$ can be computed by turning the approximate result \eqref{eq:final} into an equation using a multiplicative numerical constant $K_0$ in the right hand side and subsequently inserting literature values for the properties of HFE-7000 at $T_0 = 22$ $^\circ$C (see SI).  
For the reasonable value $K_0 = 1.6$ this results in $U_0 \approx 0.52$ m/s, in fair agreement with our experimental findings. This is quite remarkable noting the number of approximations that enter the above derivation. 

\subsection*{Numerical simulations}

To further support our findings, we have performed a series of simulations using an in-house boundary integral (BI) method, based on the code first described in  \cite{Oguz1993DynamicsNeedle}. The method has proven to be reliable for solving potential free surface flows and has been previously adapted to simulate drop impact \cite{Bouwhuis2012MaximalImpact}. For the numerical simulations, the flow inside the droplet is assumed to be a potential flow and the droplet shape to be axisymmetrical. This system is solved with the BI method to yield the time evolution of the droplet contour, where it is coupled to a compressible and condensable lubrication flow in the vapor layer. This is in contrast to the work in \cite{Bouwhuis2012MaximalImpact}, where for the non-condensable air layer it had been coupled to an incompressible lubrication flow, and to \cite{hicks2018lng}, where a different phase change model was used. More details of the simulation technique and the numerical setup can be found in the SI.

\begin{figure}
    \centering
    \includegraphics[width=0.9\linewidth]{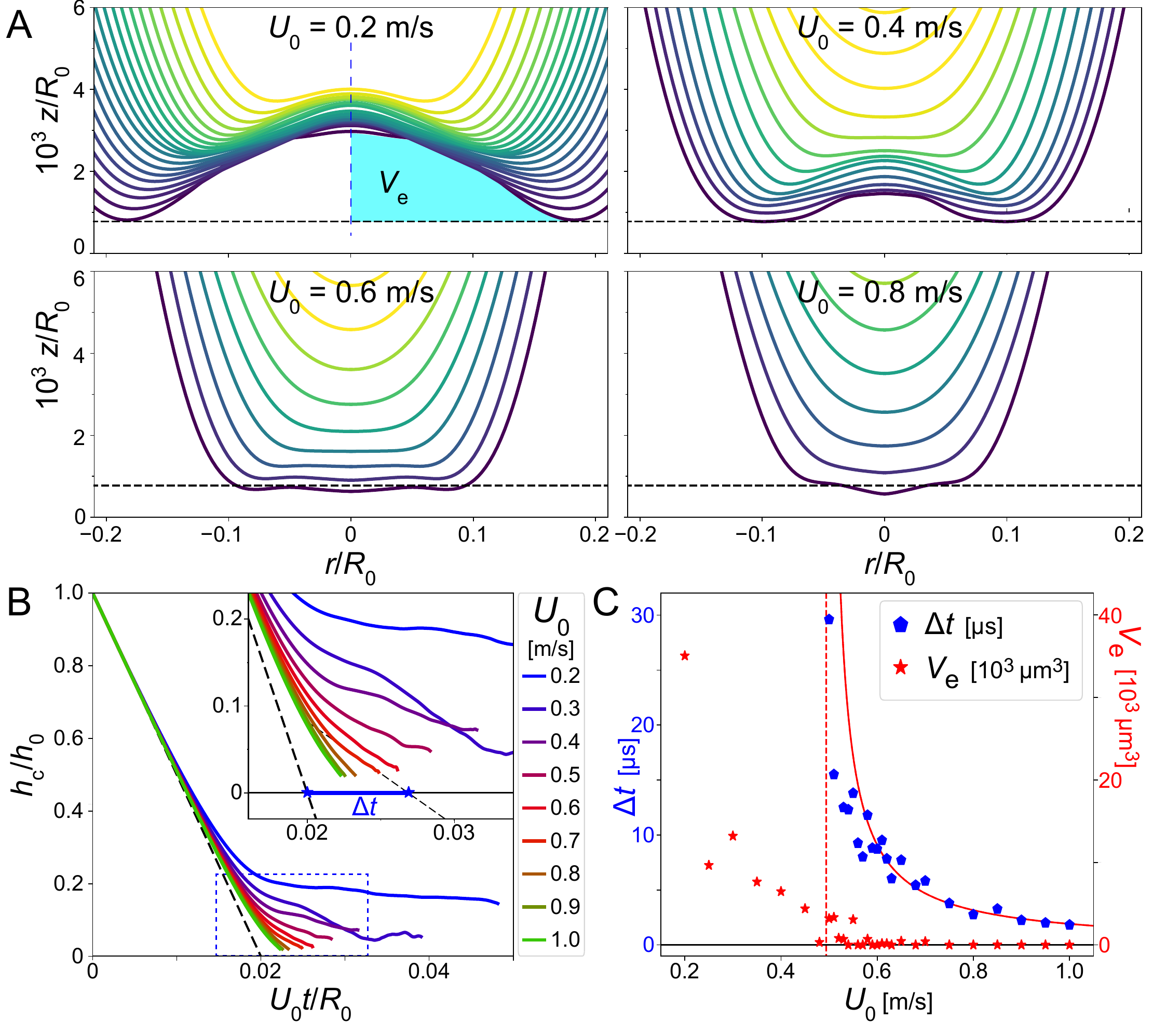}
    \caption{ \label{fig:profile_evo} (\textit{A}) Time evolution of the axisymmetric and normalized profile $h/R_0$ as a function of the normalized radial coordinate $r/R_0$ for four boundary integral simulations of a HFE-7000 droplet impacting at different speeds $U_0 = 0.2$, $0.4$, $0.6$, $0.8$ m/s.
The horizontal black dashed lines at $h = 455$ nm indicate the height at which a droplet would enter the evanescent wave of the light source, and would thus show up in a TIR experiment. Color indicates time progression from yellow to dark blue.  
(\textit{B}) Time evolution of the center height $h_c$, normalized with the initial height $h_0$, for various impact velocities $U_0$ between $0.2$ m/s and $1.0$ m/s (blue to yellow).
The straight dashed black line represents the time evolution of the position of the bottom of a moving sphere in the absence of vapor. Inset: Zoom into the blue dashed rectangular area.  
(\textit{C}) The center touchdown time $\Delta t$ (blue pentagons), plotted as a function of the impact velocity $U_0$, where the continuous red curve is a hyperbolic fit to the data and the dashed red vertical line represents its asymptote. The red stars represent the entrapped vapor volume $V_\text{e}$ indicated in (\textit{A}).
}
\end{figure}

We performed numerical simulations for similar conditions as the experimental ones reported in Fig.~\ref{fig:isothermal_evo}\textit{A}, starting the simulation from an (arbitrary) height $h_0$, large enough to not influence results. More specifically, in Fig.~\ref{fig:profile_evo}\textit{A} we extract the shape of the lower portion of an impacting droplet from the simulations. For each case, subsequent profiles are shown in time intervals of $1.33 \cdot10^{-3} R_0/U_0$, such that, far from the substrate, successive profiles for different $U_0$ would be equidistant, and the last profile corresponds to just before the simulation has stopped due to overlapping with the substrate.

For a small impact velocity ($U_0 = 0.2$ m/s in Fig.~\ref{fig:profile_evo}\textit{A}) a clear dimple is formed, which will entrap vapor upon touchdown: Starting from the first (upper) profile, which is still monotonically decreasing towards the center ($r/R_0 = 0$), one observes that the surface in the center is halted and only moves downward slowly, forming a dimple in the center, which subsequently will lead to the entrapment of a vapor bubble. After the last profile shown, the liquid will make contact with the substrate at the minima in the profile. 
For the next impact velocity, at $U_0 = 0.4$ m/s, we see a similar profile sequence as for $U_0 = 0.2$ m/s, with some subtle differences: The dimple formation happens closer to the surface, and towards the end the central part of the profile moves downward slightly faster. 

At $U_0 = 0.6$ m/s, a qualitative difference appears in the dynamics of the bottom surface. From the profile sequence it becomes clear that also in this case the profile in the center is decelerated, but unlike what happened in the previous case, it continues to move downwards, and the bottom is nearly flat such that, even if some vapor is entrapped upon impact, it is likely to be very short lived: The subsequent build-up of pressure in the entrapped bubble will be able to quickly condense whatever vapor is left inside it. Finally, at the highest impact velocity ($U_0 = 0.8$ m/s in Fig.~\ref{fig:profile_evo}\textit{A}) there is no sign of dimple formation at all: The droplet falls through the vapor layer as if no intermediate phase existed and hits the substrate with its center, without entrapment of a vapor bubble. Note that the spacing between profiles becomes larger with increasing speed, indicating considerably less deceleration for large $U_0$. All these findings are consistent with the experimental data presented in Fig.~\ref{fig:isothermal_evo}\textit{A}. In the SI (Fig.~S3) we compare the profiles of Fig.~\ref{fig:profile_evo}\textit{A} to those of a non-condensable gas with the same properties.

Summarizing, in both experiment and simulation we find similar behavior: For small impact velocities a dimple is formed that leads to the entrapment of a vapor bubble, at intermediate velocities ($U_0 \approx 0.5$ m/s) the dimple shape becomes a lot less prominent, consistent with the speed where we no longer observe entrapment in the experiments. For the highest impact velocity shown, neither a dimple shape, nor an entrapped vapor bubble can be discerned at all.

We now proceed to use the simulations to determine the transition impact velocity $U_{0,\text{thr}}$ from vapor entrapment to immediate touchdown of the liquid surface. To quantify $U_{0,\text{thr}}$, we determined the time difference $\Delta t$ between the moment of impact of the bottom of a sphere approaching the substrate in the absence of an intermediate phase and the impact time of the centre of the bottom surface of a simulated drop impact. To obtain the latter, we have tracked the center height $h_c$, which is the distance from the substrate to the bottom liquid surface at the center, i.e., at $r=0$, for simulations with impact velocities $0.2 \leq U_0 \leq 1$ m/s, as shown in Fig.~\ref{fig:profile_evo}\textit{B}. As already mentioned, the simulations will stop a little before contact, such that we do not possess numerical data all the way to touchdown and linearly extrapolate the dimple height to determine $\Delta t$ as indicated in the inset of Fig.~\ref{fig:profile_evo}\textit{B}. 

The result is found in Fig.~\ref{fig:profile_evo}\textit{C} (blue pentagons). For large $U_0$, central touchdown times are small ($\Delta t < 10$ $\mu$s), and would correspond to cases where in experiment no vapor bubble is entrapped. Approaching a threshold 
velocity, $\Delta t$ starts to diverge, and below $U_0\approx0.5$ m/s $\Delta t$ becomes very large.
This corresponds to the entrapment regime.
The results are fitted to a hyperbolic function, from which the vertical asymptote is determined as $U_{0,\text{thr}}=0.50 \pm 0.02$ m/s. This value is close to the transition value observed from bubble entrapment to touchdown in the experiments. The red stars represent the entrapped vapor volume $V_\text{e}$, estimated as the volume that is located above and within the minimum of the last profile (the lightblue region indicated in Fig.~\ref{fig:profile_evo}\textit{A}, $U_0 = 0.2$ m/s). Note that for a non-condensable gas the entrapped volume gradually decreases with $U_0$ \cite{Bouwhuis2012MaximalImpact}, but not abruptly. This is discussed more extensively in the SI (Fig.~S4).

\subsection*{Impacts on a moderately superheated surface}

To complement the isothermal experiments, we have performed experiments with a significantly heated substrate, which in principle should counteract the condensation. This is reminiscent of the well-known Leidenfrost effect \cite{Quere2013LeidenfrostDynamics}, where a droplet is made to hover over a strongly heated substrate levitated by its own vapor. The result is shown in Fig.~\ref{fig:isothermal_evo}\textit{B}, for droplets impacting at $U_0 \approx 0.7$ m/s. For superheat $\Delta T = T_s - T_0 = 2.4$ K, the entrapment is completely suppressed, as based on our isothermal experiments would be expected for this impact velocity. 
However, as soon as we increase the substrate temperature the droplet will start to entrap vapor again. For four degrees higher substrate temperature ($\Delta T = 6.4$ K) the lifetime of the vapor bubble is extremely brief (less than 10 $\mu s$), the initial entrapment radius small, and moreover only visible in the first frame.
At $\Delta T = 11.3$ K, the entrapped vapor bubble remains visible in the first two frames and further incrementing $T_s$ to $\Delta T = 17.6$ K we fully recover vapor entrapment: An axisymmetric wetted ring is found in the first recorded frame and the inner and outer contact lines expand as was previously observed in the isothermal low impact velocity experiments, where the entrapped vapor bubble remains visible up to the last frame shown in the figure. 

\begin{figure}
\begin{center}
\includegraphics[width=0.9\linewidth]{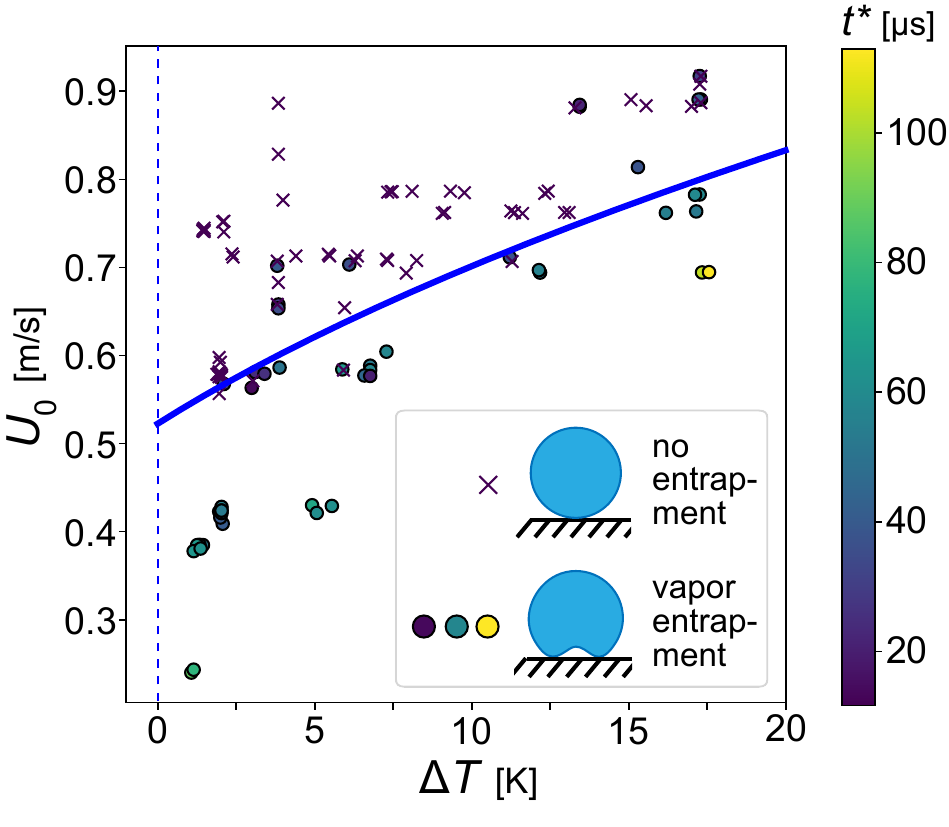}
\caption{\label{fig:phase} Experimental phase diagram tracing the boundary between droplet impacts with (colored circles) and without (crosses) vapor entrapment. On the vertical axis the impact velocity $U_0$ is plotted, whereas on the horizontal axis we find the temperature difference $\Delta T$ of that of the heated substrate and the equilibrium temperature of the surrounding vapor. The color of the circles denotes the bubble lifetime $t^*$, as indicated by the color bar. The blue curve indicates the phase boundary predicted from the model.}
\end{center}
\end{figure}

In summary, the process by which entrapment of a vapor bubble below a droplet impacting at higher velocities is suppressed, can be reversed by heating the substrate, frustrating the condensation that occurs at the vapor-liquid interface. The phase diagram in Fig.~\ref{fig:phase} traces the transition from vapor bubble entrapment to its suppression 
in the parameter space formed by the substrate superheat $\Delta T$ 
and the droplet impact velocity $U_0$. Black crosses denote impacts without entrapment and circles indicate that a vapor bubble was entrapped, where their color represents the lifetime $t^*$ of the vapor bubble, changing from dark blue to bright yellow for increasing lifetime. For nearly isothermal impacts, no entrapment is recorded above $U_{0,\text{thr}} \approx 0.6$ m/s, in good agreement with the numerical simulations we conducted. When the superheat $\Delta T$ is increasing, this transition point keeps shifting to larger velocities. Notably, the entrapment lifetime along the transition line remains small, whereas bubbles corresponding to points further away from the transition line live longer.

Inspired by the analysis of the ordinary \cite{Quere2013LeidenfrostDynamics} and dynamic \cite{gordillo2022initial} Leidenfrost phenomenon, the superheat $\Delta T$ can be included by adding a parasitic linear heat conduction term through the vapor layer, $Sk_V (T_s - T_V)/h \approx Sk_V\Delta T/h$ to the left hand side of \eqref{eq:massheatbalance}, where $h = h_d$ is again estimated from the dimple condition \eqref{eq:dimpleStokes}. Using a second multiplicative numerical constant $Q_0$ in front of the parasitic heat conduction term, the condition $U_{cond}/U_0 = 1$ for $Q_0 = 1.35$ leads to the blue line in Fig.~\ref{fig:phase}, which nicely demarcates the regions with and without vapor bubble entrapment in Fig.~\ref{fig:phase}. Details of this calculation can be found in the SI. 

Note that the transition from vapor bubble entrapment to no entrapment is gradual (second order) rather than abrupt (first order). This, in general, makes it more difficult to find the phase boundary in an experiment. In addition, the divergence observed in the touchdown time $\Delta t$ observed in the numerical simulation is an artificial one since condensation in principle continues and eventually the vapor bubble would disappear also in the simulation, which is why the dimple height bends back towards zero for the smallest impact speeds in Fig.~\ref{fig:profile_evo}\textit{B}, leading to large but finite touchdown times. 
In the experiment there is the added effect (at least for small or zero superheat) that HFE-7000 is observed to be very wetting, which will lead to entrainment of the vapor bubble into the inside of the droplet.

\subsection*{Conclusions}

In conclusion, we have performed impact experiments of droplets under boiling liquid conditions, i.e., of a liquid that is in thermal equilibrium with its own vapor. Whereas, for substrate temperatures close to that of the surroundings, at low impact speeds a vapor bubble is entrapped --similar to the entrapment of an air bubble for droplet impact under atmospheric conditions--, we observe that vapor bubble entrapment is suppressed completely at sufficiently high impact speeds. Heating the substrate, we can counteract this effect: If we start with a sufficiently high impact speed for which no vapor bubble is entrapped, we can by progressively heating the substrate reach a situation where vapor entrapment reappears. 

We explain this behavior from condensation at the vapor-liquid interface: If the pressure build-up below the droplet is sufficiently high to cause the vapor to condense, at higher impact velocities we arrive in a situation where condensation becomes fast enough to consume the entire intermediate vapor layer during impact, and the droplet hits the surface without the cushioning action of the gas layer that is present during impact under atmospheric conditions. 

We create an approximate model based on scaling arguments to explain our findings, which is in good agreement with our experiments. The argument starts from the same mass-energy balance approximation at the liquid-vapor interface we also used successfully in explaining the strongly increased pressures observed during wave \cite{ezeta2025large} and disc \cite{fan2025impact} impact in a boiling liquid, which provides additional confidence in the correctness of this approach. 

In addition, we devised a numerical simulation consisting of a boundary integral method for the impacting droplet, coupled to a compressible lubrication model including phase change for the vapor layer that provides insight into the vapor dynamics below the impacting droplet. Since gas compressibility effects in droplet impact are important and constitute an active research area \cite{hicks2013liquid,sprittles2024gas}, it is good to stress that we are in an impact regime where compressibility effects are expected to be minor \cite{Mandre2009PrecursorsSurface}, as discussed in the SI. However, phase change (condensation) can only be reliably incorporated into a compressible lubrication model.

Our research is of consequence for the transport of cryogenic liquids such as LNG and LH2, where the suppression of vapor bubble entrapment is likely to go hand in hand with increased impact pressures. The models introduced in this work may be translated to the conditions in those applications to verify under which circumstances phase change may become a non-negligible effect and needs to be included into the engineering of cryogenic liquid carriers on an industrial scale. More specifically, using our theoretical results in estimating when the rate of condensation exceeds the impact speed of a sloshing wave crest, will make it possible to identify and avoid situations in which the protective cushioning effect of the intermediate vapor phase is cancelled by condensation. The latter will increase the impact load, potentially threatening the structural integrity of the transporting containers.

\bibliographystyle{prsty_withtitle}
\bibliography{references_mendeley_ext}

\end{document}


\begin{center}
    {\LARGE \bfseries Supporting Material}\\
     \vspace{1em}
     {\Large \bfseries Impact of boiling liquid droplets:\\ 
    Vapor entrapment suppression \par}
    \vspace{1em}
    Bernardo Palacios-Mu\~{n}iz, Edgar Ortega-Roano, Yee Li (Ellis) Fan,\\
    Nayoung Kim, and Devaraj van der Meer, \par
    \vspace{0.5em}
    {\small\it
      Physics of Fluids Group, Max Planck Center Twente for Complex Fluid Dynamics,\\ University of Twente, 7500 AE Enschede, The Netherlands.
    }\par
 \end{center}

This Supporting Text contains six different Sections. In Section 1 we discuss the experimental setup in greater detail than was done in the main text. The second Section provides additional information on the transport properties for HFE-7000 that are used in the main text and how they were obtained from the literature. The very small Section 3 discusses the capillary bubble entrapment regime for HFE-7000, whereas the fourth Section justifies our choice for the parameter $\epsilon$ which estimates the growth time of the thermal boundary layer in the liquid. Section 5 provides details on the numerical method used to compute the vapor layer dynamics including phase change and compares the results presented in the main text to those in a hypothetical non-condensable gas which has the same properties as the HFE-7000 vapor. Finally, Section 6 provides some additional calculations extending the simplified model from the main text to include the substrate superheat $\Delta T$.    

\section{Details of the experimental setup}

As discussed in the main text, in order to generate droplets in equilibrium with their own vapor and record their impact onto a substrate, we have developed a highly versatile setup, using HFE-7000 (Heptafluoropropyl-methyl-ether, also known by the brand name Novec 7000) as a working fluid. This fluid was selected due to its low boiling point ($\sim34$ $^\circ$C), which in turn ensures an appreciable equilibrium vapor pressure ($P_{V,0} = 0.57$ bar) at room temperature, which is both advantageous for performing experiments, and reasonably close to conditions found in cryogenic liquids. A  sketch of the setup can be found in Fig.~\ref{fig:setup_sketchAPP}\textit{A} (which is identical to Fig. 1 in the main text, but contains a more extensive description of its various parts in the caption) and this Section contains a detailed description of its different components.

\subsection*{Creating a boiling liquid}
When preparing an experiment, the first step is creating an environment saturated with pure vapor (at its vapor pressure) in which we can create droplets of liquid. These conditions are established in a hermetically closed chamber with a wall temperature $T_0$ set by cooling/heating channels in the wall of the setup and a thermal circulator (TC). For the vapor to be as free of humidity as possible, the chamber is then vacated by means of a pump (VP) to an absolute pressure $P_{0}<5$ mbar and flushed with nitrogen (N2). This procedure is repeated at least $3$ times and the chamber is vacated one last time to $P_0 < 3$ mbar. At this stage, liquid HFE-7000 is allowed to flow into a reservoir inside the chamber and, due to the low pressure, the liquid will evaporate immediately and start saturating the chamber with vapor. Once the setup is saturated, liquid HFE-7000 will continue to flow into the reservoir. Hereby, enough liquid is allowed into the reservoir to ensure the chamber remains saturated with vapor during the entire experiment. 
Due to the fast evaporation the temperature will drop (especially close to the reservoir), however, with time the stored liquid HFE-7000 will heat up from its surroundings and continue evaporating. In case the temperature drop is too far from our desired set point, we can use a resistive heater which is attached to the reservoir to accelerate this process. 
During the procedure, we monitor the pressure $P_{0}$ and temperature $T_{0}$ in the setup and wait for steady state readings, indicating that thermal equilibrium is established. Our measured steady state vapor pressure is in very good agreement with that measured in \cite{Ohta2001Liquid-PhaseEther}, for the same liquid.

\subsection*{Droplet generation and impact}
Additional to the liquid HFE-7000 in the main reservoir, some extra liquid is allowed into a smaller reservoir located higher in the setup and, from it, a capillary tubing line is connected to a needle (N) set at a lower height than the liquid level. Droplets are then simply generated by gravity. In between the needle and reservoir there is an on/off valve (V4) which can be controlled from outside of the setup to stop the flow. Once the whole system is at equilibrium, liquid is allowed to flow to the needle.   
Typically, some vapor will be initially entrapped in the needle and droplets will be ejected with irregular shapes from the tip, so droplets fall onto a catcher until uniform droplets with an initial radius $R_0 \approx 0.6$ mm are formed at an adequate rate. The catcher can be moved remotely such that a single droplet is allowed to pass and impact a substrate, consisting of a sapphire prism of high thermal conductivity. A resistive heater is mounted to the side of the prism, such that the temperature of the substrate, which is probed continuously with a PT-100 sensor, is kept within $\pm 0.1$ K of a set temperature $T_s$ between $T_0$ and $40$ $^\circ$C. Finally, the height between needle and substrate can be varied to obtain an impact velocity $U_0$ between $0.2$ m/s and $1.5$ m/s. 

\subsection*{Imaging}
We image the impact from side and bottom views. Side views are recorded at 16,000 fps with a high speed camera (CAM1, Photron Nova S12 or S16) coupled to a long distance microscope (LDM, Navitar 12x system). This view uses backlighting with a white (6,500 K) and well collimated light source (CLS1) to create a sharp image of the falling droplet and is used to measure the initial droplet radius $R_0$ and impact velocity $U_0$. The imaging system is calibrated/corrected with a resolution/distortion test target set in the same focal plane as the falling droplets.

The bottom views are captured with frustrated total internal reflection (TIR) imaging, which consists in illuminating the top surface of an inverted prism (P) at an angle greater than the critical angle for total internal reflection on the solid-gas/vapor interface, but smaller than that for the solid-liquid interface. If a droplet is within the evanescent wave distance from the surface, then light is no longer totally reflected and upon contact light can fully transmit into the droplet. So the light intensity at any point of the surface can be used as a measure of the droplet's proximity to that point \cite{Kolinski2014Lift-OffSurface,Shirota2017MeasuringFTIR}. This technique has been most prominently used to not only distinguish wetting of the surface, but even to quantitatively measure the nanometric film thickness of the gas layer underneath impacting droplets \cite{Kolinski2014Lift-OffSurface,VanLimbeek2016,Chantelot2021DropContact}. However, to do the latter requires the use of coherent and polarized light and complex calibration procedures. Due to space constraints inside the setup, it is impossible to fit the elements required to quantify the thickness of a gas or vapor film, but we can very clearly distinguish wet (or very nearly wet) and dry zones of the surface. To do so, we collimate a monochromatic ($455$ nm, nominal) LED light source (CLS2) and illuminate a sapphire prism such that it totally internally reflects in the top surface. Images of the surface are recorded with an infinity corrected microscope objective (Mitutoyo 5x), tube lens (TL, Thorlabs) and high speed camera (CAM2, Photron Nova S16) at frame rates between $96,000$ and $200,000$ fps, 
which are sufficient to study the early impact dynamics of the droplets. Because of the short wavelength of the light, any reduction in intensity from the totally reflected background implies that the liquid is within less than half a micron of the solid surface.

\subsection*{Description of a TIR sequence from a typical impact}
\begin{figure}
\centering
\includegraphics[width=0.95\linewidth]{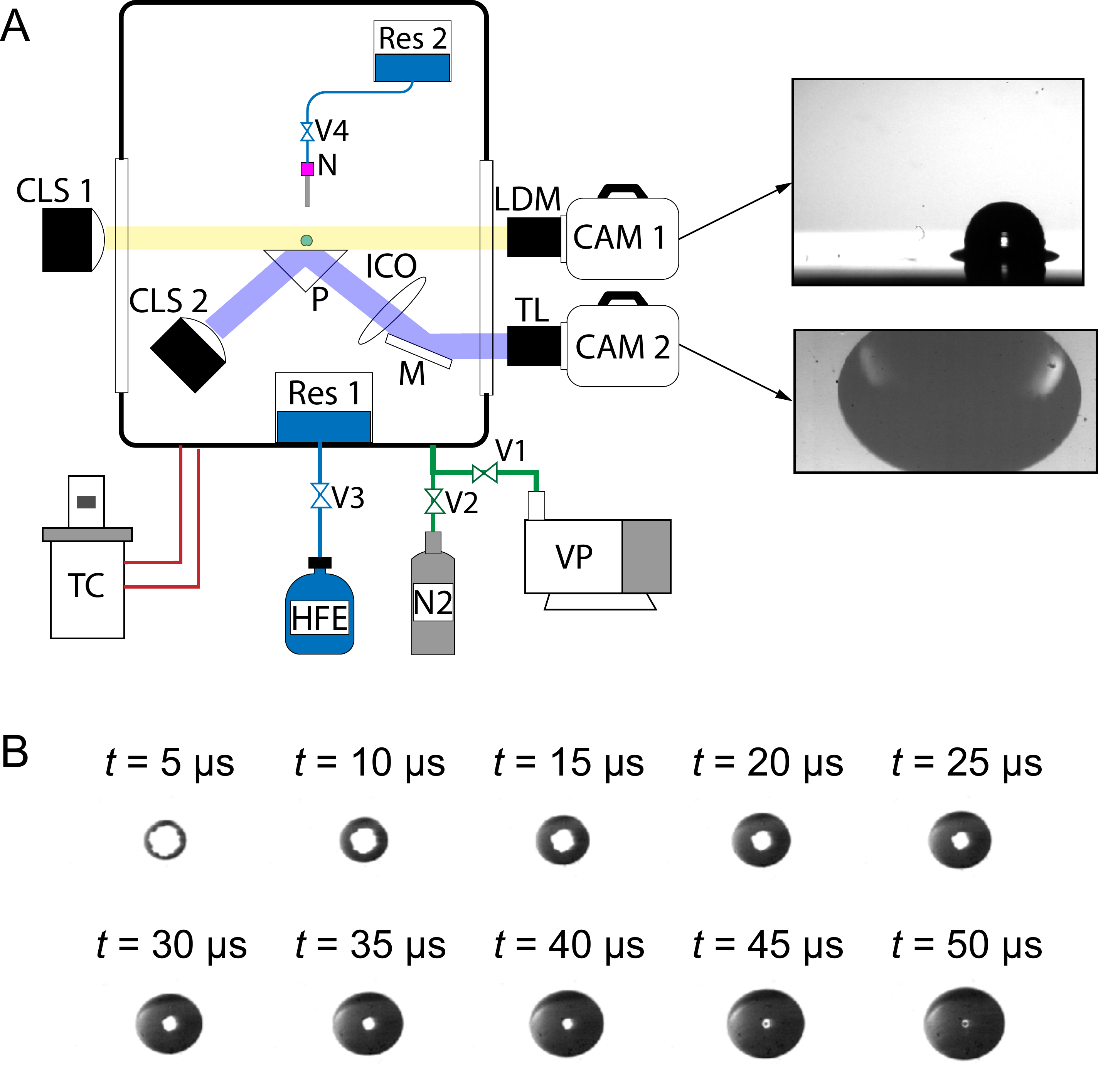}
\caption{\label{fig:setup_sketchAPP} (\textit{A}) Sketch of the experimental setup with as main elements the closed, temperature-controlled chamber containing HFE-7000 liquid (Res1) in equilibrium with its vapor, the droplet generated from the second reservoir (Res2), and the prism (P) used for frustrated total internal reflection imaging (TIR), the vacuum pump (VP), and the two high-speed cameras (CAM1,2) used for imaging. More specifically, a  
closed chamber with ambient temperature $T_{0}$ (set by a thermal circulator TC) is flushed with nitrogen and vacated with a vacuum pump VP, via valves V1 and V2. HFE-7000 is allowed into the chamber via valve V3 and saturates the atmosphere inside the chamber with vapor.\\ \emph{[Caption continued on next page:]}}
\end{figure}

Fig.~\ref{fig:setup_sketchAPP}\textit{B} shows a typical (near-)isothermal experiment for a droplet impacting a substrate at a small impact velocity of $U_0 = 0.38$ m/s. The frames are a TIR sequence where wet (dark) and dry (light) zones are well distinguished. On the first frame, a dark ring appears over the background; the ring is the surface of the droplet as it just starts making contact with the substrate and the light part within the ring is the dimple which has formed under the droplet and contains entrapped vapor. As the droplet further contacts and expands on the surface, the ring expands both in the inner and outer directions. It is also noteworthy that, for these particular conditions, the last two frames of \ref{fig:setup_sketchAPP}\textit{B} show a small wet point at the very middle of the cavity. This point suggests a downward jet has been ejected towards the sapphire surface similar to those observed in \cite{Zhang2022ImpactSurfaces} and  \cite{Lee2020DownwardDrop}.

The closure of the ring implies that the dimple's neck (which is that part of the liquid closest to the surface) is changing its radial position. This is atypical to happen before impact, where for isothermal impacts in air over a smooth surface the neck has been observed to be usually fixed in position \cite{Kolinski2014Lift-OffSurface}. This therefore indicates that here the liquid is actually wetting the surface and could suggest entrapment of a bubble by the liquid as described in \cite{THORODDSEN2005TheSurface}, where the bubble moves upward inside the liquid. This in turn could well be connected to the excellent wetting properties of HFE-7000 on sapphire in particular, but otherwise for almost any substrate.
In Fig.~\ref{fig:extraimages} we show additional snapshots for each of the sequences presented in Fig.~2\textit{A} in the main text. 

\section{Transport properties of HFE-7000}

In this Section, we clarify the way we obtain the temperature-dependent physical properties of HFE-7000 from several references \cite{20213MFluid,widiatmo2001equations,ohta2001liquid,rausch2015density,perkins2022measurement,aminian2022ideal}. Table \ref{table:NovecProp} summarizes the temperature-dependent properties of HFE-7000 for $T_0 = 22$ $^\circ$C, that are used for the computation of $U_{cond}/U_0$ using Eq. (9) in the main text, for the parameters used in the numerical simulation (see Section 5 of this SI Appendix) and for the computation of the phase boundary in Fig. 5 of the main text (see also this SI Appendix, Section 6).

The saturation pressure $P_V$ for HFE-7000 at different (absolute) saturation temperatures $T_V$ is computed based on the empirical and experimentally confirmed correlation provided in \cite{ohta2001liquid}:
\begin{equation} \label{eq:supp pv0}
    \text{ln} \ P_\text{r} = \frac{T_\text{c}}{T_V} \left( a_1\tau +a_2\tau^{1.5} +a_3\tau^3 + a_4\tau^6 \right)
\end{equation}
where $P_\text{r} = P_V/P_\text{c}$ and $\tau = 1 - T_V/T_\text{c}$. Here, subscript c indicates the critical temperature $T_\text{c}$ and critical pressure $P_\text{c}$. For HFE-7000, the values are $P_\text{c} = 2476$ kPa, $T_\text{c} = 437.7$ K, $a_1 = -7.951$ , $a_2 = 1.510$, $a_3 = -4.481$ and $a_4 = -20.835$. This correlation is used as the reference, rather than the one provided in the 3M data sheet \cite{20213MFluid} because repetitive measurements of the saturation pressure of HFE-7000 at various temperatures consistently coincide with the one provided in \cite{ohta2001liquid} (see~\cite{muniz2024impact}). With this, the equilibrium pressure $P_{V,0} \equiv P_V(T_{0})$ is computed with \eqref{eq:supp pv0}. Using the ideal gas law, we calculate the equilibrium vapor density $\rho_{V,0}$ as
\begin{equation}
    \rho_{V,0} = \frac{P_{V,0}}{R_\text{s}T_{0}}\,\,,
\end{equation}
where the specific gas constant $R_\text{s} = R_\text{u}/M$, with $R_\text{u} = 8.314$ J/mol K the universal gas constant, $M = 0.200$ kg/mol the molar mass of HFE-7000, and $T_0$ expressed in the Kelvin scale. 

\begin{table}[h!]
\centering
\begin{tabular}{|c| c| c| c| c| c| c|} 
\hline
 $T_{0}$ & 
 $P_{V,0}$ & 
 $\rho_{V,0}$ & 
 $\rho_{L}$ &
 $c_{p,V}$ &
 $c_{p,L}$ &
 $k_V$ \\ [0.5ex] 
 
 \hline 
[$^{\circ}\text{C}$] & 
[bar] & 
[kg/m$^3$] & 
[kg/m$^3$] & 
[kJ/kg K] & 
[kJ/kg K] & 
[W/m$\cdot$K] 
\\ [0.5ex] 

 \hline
$22$ &	
$0.618$ &	
$5.04$ & 
$1423$ & 
$0.884$ &
$1.17$ &
$0.0117$
\\ [0.5ex] 

 \hline\hline
 $k_L$ &
 $\alpha_V$ &
 $\alpha_L$ &
 $\mu_V$ & 
 $L$ &  
 $\sigma$ &
 $\beta$
 \\ [0.5ex] 

 \hline
 [W/m$\cdot$K] &
[mm$^2$/s] &  
[mm$^2$/s] &  
[$\mu$Pa$\cdot$s]  & 
[kJ/kg] & 
[mN/m] &
[-] \\ [0.5ex] 

\hline
$0.0755$ &
$2.62$ &
$0.0455$ &
$11.7$ &	
$138$ &	 
$12.1$ &
$0.0888$
\\ [1ex] 
\hline
\end{tabular}
\caption{\textmd{Properties of HFE-7000 relevant to the calculations in the simplified model and the numerical simulations at $T_0 = 22$ $^\circ$C. Subscript V is for the vapor phase while subscript L stands for the liquid phase.}}
\label{table:NovecProp}
\end{table}

The properties of liquid HFE-7000 are mainly derived from \cite{aminian2022ideal}, which comes with a table containing data of the thermodynamic properties along the saturation lines for HFE-7000 obtained using the Peng-Robinson equation of state, which is a cubic equation of state. We interpolate the provided data with cubic interpolation to obtain the latent heat of vaporization $L$, the specific heat capacity $c_{p,L}$ at constant pressure, and the density $\rho_L$ in the liquid phase, corresponding to our experimental equilibrium temperature $T_0$. Meanwhile, the thermal conductivity $k_L$ of liquid HFE-7000 is calculated with the formula provided in the 3M data sheet \cite{20213MFluid}
\begin{equation}
    k_L = 0.0789 - 0.000196T_0\,\,,\label{eq:kL}
\end{equation}
which coincides well with the measurements provided in \cite{perkins2022measurement}. Note that in \eqref{eq:kL} $T_0$ is expressed in $^{\circ}$C. The thermal diffusivity $\alpha_L$ of liquid HFE-7000 can then be calculated as:
\begin{equation}\label{eq:alpha}
    \alpha_L = \frac{k_L}{\rho_Lc_{p,L}}
\end{equation}
 
The thermal conductivity of HFE-7000 vapor is provided in \cite{perkins2022measurement} with the empirical relationship:
\begin{equation}
    k_V = \frac{ A_0 + A_1\tau_V + A_2\tau_V^2 }{ a_0 + a_1\tau_V + a_2\tau_V^2 + a_3\tau_V^3 + a_4\tau_V^4 } 
\end{equation}
where $\tau_V = T_V/T_\text{c}$, $a_i$ are constants of values $a_0$ = 34.1702, $a_1$ = -49.3874, $a_2$ = 44.4355, $a_3$ = -11.058, $a_4$ = 1.0, and $A_0$ = 0.122535 W/m.K, $A_1$ = -0.29099 W/m.K and $A_2$ = 0.621481 W/m.K are the dilute gas, residual and critical thermal conductivity, respectively. In addition, we provide the dynamic viscosity $\mu_V$ of the vapor phase, obtained by cubic interpolation of the data provided in \cite{rausch2015density} (see Table 4 therein). 

In general, at room temperature, the momentum (viscosity) and thermal (conductivity/diffusivity) transport properties of HFE-7000 in both liquid and vapor phases are slightly smaller than that of water and air, but still of the same order of magnitude, except for the thermal conductivity of liquid HFE-7000 which is an order of magnitude smaller than that of water. It is also good to note that the density ratio between the vapor and liquid $\rho_{V,0}/\rho_L$ for the equilibrium temperature settings in our experiment is larger than the density ratio between air and water, which is around $0.0012$.

The small value of the surface tension $\sigma$ together with the large liquid density $\rho_L$ leads to a small capillary length $l_c = \sqrt{\sigma/(\rho_Lg)} \approx 0.931$ mm, which limits the length scales for which one may obtain spherical droplets.

\section{Inertial versus capillary bubble entrapment regimes}

In describing the entrapment of a non-condensable gas bubble below an impacting droplet, in \cite{Bouwhuis2012MaximalImpact} a distinction is made between the inertial regime, which is the regime that we are describing in the main text and for which the Stokes scaling holds, and the capillary regime, where surface tension is responsible for the shape of the entrapped bubble. The crossover velocity between the inertial and the capillary regime is provided as \cite{Bouwhuis2012MaximalImpact}
\begin{equation}
U_\text{cr} \sim \frac{\mu_g^{1/7}\sigma^{3/7}}{\rho_L^{4/7}R_0^{4/7}}\,\,,    
\end{equation}
where $\sigma$ is the surface tension. Inserting the typical values for HFE-7000 (with $\mu_g = \mu_V$) gives $U_\text{cr}\sim 0.0326$ m/s, which is almost an order of magnitude smaller than the smallest experimental impact velocities reported in this work. This confirms that the inertial regime is indeed the correct choice to describe our findings.

\section{Dimensionless growth time of the thermal\\ boundary layer in the liquid}

The impacting droplet and (unheated) substrate are initially in thermal equilibrium, so there is no temperature difference between them. Therefore a temperature difference $\Delta T_V$ between the vapor and the droplet only starts to occur as soon as the lubrication pressure in the vapor layer starts to rise, which only happens when the vapor layer becomes slender, i.e., $h/R_0 \ll 1$, where $h$ is the distance of the (undisturbed) droplet of radius $R_0$ to the substrate.  In terms of the time $\tau = h/U_0$ remaining until impact of the (undisturbed) droplet in vacuum, we obtain $\epsilon = U_0\tau/R_0 = h/R_0 \ll 1$. This leads to a ``common sense'' order-of-magnitude estimate for $\epsilon$ of $10^{-2}$. In the following we will take a more quantitative approach.

Since both pressure and temperature difference will start to rise monotonically, the expression $\delta_\text{th} \approx\sqrt{\pi\alpha_L\tau}$ can only be considered to be an estimate of the true boundary layer thickness, since the former would correspond to the situation that, at time $\tau$, the temperature would suddenly rise to a constant value and remain there until the dimple is formed. Therefore, to estimate $\tau$ a bit more quantitatively, we need to estimate the moment at which the pressure in the vapor layer rises appreciably, which we can do by estimating the maximum pressure $\Delta P_\text{max}$ in the lubrication layer. Estimating the typical Stokes number as $\mathit{St} = \rho_L U_0 R_0/\mu_V \approx 1423 \cdot 0.4 \cdot 0.0006 / (1.17\cdot10^{-5}) = 2.9\cdot10^{4}$ (where we used $U_0 = 0.4$ m/s as a typical value), we can use the scaling relation from \cite{Bouwhuis2012MaximalImpact} to estimate the dimple height as $h_d/R_0 \approx 3.0 \,\text{St}^{-2/3} \approx 3.2\cdot10^{-3}$. With Eq. (1) from the main text we can subsequently estimate the maximum pressure as $\Delta P_\text{max} \sim \mu_VU_0/R_0 (h_d/R_0)^{-2}$. (Note that this estimate corresponds to the worst case scenario of a non-condensable gas/vapor.) Now if we (rather arbitrarily) define the pressure at which the effect becomes appreciable at 1\% of $\Delta P_\text{max}$, this directly  leads $h/R_0 \approx 10 h_d/R_0 \approx 0.032$. Now, finally, $\epsilon = U_0\tau/R_0 = h/R_0 \approx 0.03$.

This estimate is consistent with \cite{gordillo2022initial}, where for the slightly different situation of a droplet impacting onto a (substantially) heated substrate, the boundary layer thickness was estimated as $\delta_{th}\propto \sqrt{\alpha_L \tau_m R_0/U_0}$, where $\tau_m=12.4 \mathit{St}^{-2/3}$. Using $\mathit{St} \approx 2.9\cdot10^{4}$, we obtain $\tau_m \approx 0.01$, which needs to be directly compared to our value of $\epsilon$.

From all of the above we conclude that $\epsilon \in [0.01,0.03]$. Since in our numerical simulations we saw no observable differences when starting from $U_0\tau/R_0 =0.03$ and $U_0\tau/R_0 =0.02$, and given the arbitrariness of the 1\% threshold, we have chosen the slightly more conservative value $\epsilon = 0.02$.

\section{Details of the numerical setup}

The axisymmetric simulations performed in the main text are performed using an in-house boundary integral (BI) method coupled to a compressible lubrication model for the flow in the vapor layer that includes phase change. Whereas the BI method is well established \cite{Oguz1993DynamicsNeedle}, as well as its coupling to an incompressible lubrication equation for a non-condensable gas \cite{Bouwhuis2012MaximalImpact}, treating the vapor layer using  a compressible lubrication model with phase change is new, and will therefore be discussed below. Note that a first study aimed at including phase change into a lubrication description was reported in \cite{hicks2018lng}, where however a different approach to the inclusion of phase change was taken than in this manuscript, namely using the Hertz-Knudsen relation.  

\subsection*{Lubrication model of the vapor layer including phase change}

In Fig.~\ref{fig:theory_sketch}\textit{A} we schematically show the parameters that play a role in describing the vapor layer. In the lubrication approximation, the axisymmetric Stokes equation and the no-slip boundary conditions at the substrate and the liquid-vapor interface become
\begin{equation}\label{eq:Stokes}
\frac{\partial P_V}{\partial r} = \mu_V \frac{\partial^2 u_r}{\partial z^2}  \,\,,\qquad
u_r(r,z\!=\!0,t) = 0\,\,, \qquad u_r(r,z\!=\!h,t)=V_\parallel(r,t)\,\,,
\end{equation}
where $P_V(r,t)$, $h(r,t)$ and $u_r(r,z,t)$ are the local pressure in, the height of, and the radial velocity in the vapor layer, respectively. Here, $V_\parallel$ is the radial velocity parallel to the substrate at the liquid droplet-vapor interface (as obtained from the boundary integral method), and $\mu_V$ the dynamic viscosity of the vapor. Note that changes in the temperature of the vapor are considered to be small enough for $\mu_V$ to remain constant. 
Since, due to the lubrication approximation, $P_V(r,t)$ is not a function of $z$, this equation can be solved for $z$ together with the boundary conditions and yields the velocity profile
\begin{equation}
u_r(r,z,t)=\frac{1}{2 \mu_V} \frac{\partial P_V}{\partial r} z(z-h)+ V_\parallel \frac{z}{h}\,\,.
\label{eq:flowprofile}
\end{equation}
Since we allow for phase change, we cannot directly use the continuity equation. Instead, we compute the vapor mass rate of change due to the mass that flows radially in and out of a ring-shaped control volume with boundaries at $r$ and $r\!+\!dr$, by integrating $\rho_Vu_r$ over the thickness $h$ at the two boundaries, as indicated in Fig.~\ref{fig:theory_sketch}\textit{A}. With use of \eqref{eq:flowprofile} we obtain 
\begin{equation}
\frac{d m_V}{d t} =  \frac{\pi}{6 \mu_V}\frac{\partial}{\partial r}\left(r \rho_V h^3 \frac{\partial P_V}{\partial r} \right)dr - \pi\frac{\partial}{\partial r}\left(r \rho_V h V_\parallel \right)dr\,\,,
\end{equation}
On the other hand, $dm_V/dt$ can change due to compression of the vapor and through phase change, which can be quantified as 
\begin{equation}
\frac{d m_V}{d t}= 2 \pi r \frac{\partial }{\partial t}(\rho_V h ) dr + 2 \pi r \frac{k_L}{L} \left. \frac{\partial T}{\partial z} \right \rvert_{z=h} dr\,\,.
\end{equation}
Combining the above two equations provides the lubrication equation for the vapor layer
\begin{equation} 
\frac{\partial}{\partial t} (\rho_V h) = \frac{1}{12 \mu_V}\frac{1}{r}\frac{\partial}{\partial r}\left(r \rho_V h^3 \frac{\partial P_V}{\partial r} \right)- \frac{1}{2r}\frac{\partial}{\partial r}\left(r \rho_V h V_\parallel \right)- \frac{k_L}{L} \left. \frac{\partial T}{\partial z} \right \rvert_{z=h}\,\,.
\end{equation}

Finally, we need a closure relation for the temperature gradient $\partial T/\partial z|_{z=h}$ at the liquid-vapor interface in terms of the vapor temperature $T_V$. The simplest of such closures would be the expression
\begin{equation}
\left.\frac{\partial T}{\partial z} \right|_{z=h} \approx -\frac{T_V-T_0}{\sqrt{\pi \alpha_L t}}\,\,,
\end{equation}
which would correspond to solving the heat equation in the liquid using a constant temperature $T_V$ in the vapor phase as a boundary condition, and is also the expression we have used in the simplified model in the main text. However, in reality $T_V(r,t)$ is not constant, but a function of both the radial coordinate and time. Since the vapor layer is slender, and the thermal boundary layer thickness in the liquid is expected to be small compared to the length scale on which changes in the radial coordinate take place, we may at every $r$ solve the one-dimensional heat equation
\begin{equation}
\frac{\partial \Delta T}{\partial t} + \dot{h}\frac{\partial \Delta T}{\partial z} = \alpha_L \frac{\partial^2 \Delta T}{\partial z^2}\,\,,
\end{equation}
with $\Delta T = T_L(z,t) - T_0$ and $\dot{h} = \partial h/\partial t$ (where the dependence of those quantities of the radial coordinate $r$ is suppressed for convenience). Transforming $z$ and $t$ into the so-called Lagrangian coordinates $\zeta = z - h(t)$ and $t$, this equation reduces to
\begin{equation}
\frac{\partial \Delta T}{\partial t}  = \alpha_L \frac{\partial^2 \Delta T}{\partial \zeta^2}\,\,,
\end{equation}
which, together with the boundary conditions $\Delta T(\zeta=0,t) = T_V(r,t) - T_0$, $\lim_{\zeta\to\infty} \Delta T(\zeta,t) = 0$, and the initial condition $\Delta T(\zeta,t=0) = 0$ can be solved using a Laplace transformation in time
\begin{equation}
\Theta(\zeta,s) = L[\Delta T(\zeta,t)] = \int_{t=0}^\infty \Delta T(\zeta,t) e^{-st}dt \,\,.
\end{equation}
Defining $F(s)$ as the Laplace transform of $f(t) = T_V(r,t) - T_0 = \Delta T(\zeta=0,t)$ and using standard techniques, this leads to the solution
\begin{equation}
\Theta(\zeta,s) = F(s) \exp\left(-\sqrt{\frac{s}{\alpha_L}} z \right) \,\,.\label{eq:LTF}
\end{equation}
Since the quantity of interest is $\partial \Delta T/\partial z |_{z = h} = \partial \Delta T/\partial \zeta |_{\zeta = 0}$, which, using Eq.~\eqref{eq:LTF}, has Laplace transform
\begin{equation}
\left.\frac{\partial \Theta}{\partial \zeta}\right|_{\zeta = 0} = -\sqrt{\frac{s}{\alpha_L}} F(s) = -\sqrt{\frac{1}{\alpha_L s}}\, (sF(s))\,\,,\label{eq:main}
\end{equation}
where the rewriting into the last expression looks like a triviality, but expresses the result in the product of two functions which possess an inverse Laplace transformation, where in the first we recognize the Laplace transform of $1/\sqrt{\pi \alpha_L t}$ and in the second the Laplace transform of the time derivative of $f(t) = T_V(r,t) - T_0$, i.e., of $\partial T_V/\partial t$. Using that the inverse Laplace transform of the product of two functions is their convolution in the time domain we obtain  
\begin{equation}
\left.\frac{\partial T}{\partial z} \right \rvert_{z=h} = \left.\frac{\partial \Delta T}{\partial \zeta} \right \rvert_{\zeta=0} = -\frac{1}{\sqrt{\pi \alpha_L}}\int_{\tau=0}^{t}  \frac{1}{\sqrt{ \tau}}\left.\frac{\partial T_V}{\partial t}\right|_{t-\tau}  \!\!\!\!\!\!\!d \tau \,\,,
\end{equation}
which closes the coupling to the BI method and the lubrication layer where $\partial T_V/\partial t$ is computed numerically from $T_V(t)$ obtained in earlier time steps.

Alternatively, one may define $G(s)$ as the Laplace transform of $g(t) = \partial \Delta T/\partial z |_{z = h}$, with which \eqref{eq:main} can be inverted into
\begin{equation}
F(s) = -\sqrt{\frac{\alpha_L}{s}} G(s)\,\,, 
\end{equation}
where the left hand side is the Laplace transform of $T_V(r,t)-T_0$ and the right hand side is the product of the Laplace transforms of $\sqrt{\alpha_L/(\pi t)}$ and $g(t) = \partial \Delta T/\partial z |_{z = h}$. This then leads to the convoluted expression
\begin{equation}
T_V(r,t) - T_0 = -\sqrt{\frac{\alpha_L}{\pi}} \int_{\tau = 0}^t \frac{1}{\sqrt{t-\tau}}\left.\frac{\partial T}{\partial z} \right \rvert_{z=h(r,\tau)} \!\!\!\!\!\!\!\!\!\!\!\!\!\!\!\!d \tau \,\,,
\end{equation}
which may be interpreted as a one-dimensional version of the Plesset-Zwick formula \cite{plesset1952heatdiff}, and constitutes an integral equation which can be solved numerically for $\partial T/\partial z|_{z=h}$ to close the coupling of the BI method and lubrication layer.

Finally, we stress that the state variables in the vapor layer, $P_V$, $\rho_V$ and $T_V$ are related by the Clausius-Clapeyron relation (vapor curve) and the ideal gas law, such that if one of the state variables is known, the other two can be computed as well. 

The initial condition for any given simulation is a spherical droplet with radius $R_0$, falling with a constant speed $U_0$. The bottom of the sphere is initially at a sufficient height from the substrate ($> 10 \mu$m $\approx 0.02R_0$) where the liquid surface is not yet affected by the radial vapor flow. With the BI method, a simulation will stop when the droplet surface is close enough to either intersect itself or when it almost makes contact with the substrate such that it causes the coupling with the vapor layer to break down. 

A posteriori we can verify that solving the one-dimensional heat equation is a reasonable approximation by showing that indeed the thermal boundary layer thickness is small compared to the horizontal extent of the lubrication layer. First, since the history effect caused by a rapidly increasing vapor temperature tends to partly erase the previously formed thermal boundary layer, an upper bound is found by taking the temperature to be constant, which leads to $\delta_{th} = \sqrt{\pi\alpha_L\Delta t}$, with $\alpha_L \approx 4.55\cdot10^{-08}$ m$^2$/s. The time span $\Delta t$ may be approximated by $\Delta t = \epsilon R_0/U_0$, with $\epsilon \approx 0.02$ and $R_0 \approx 0.58$ mm. Taking the worst case scenario of $U_0 = 0.2$ m/s then leads to $\delta_{th} \approx 2.9$ $\mu$m, which is much smaller than the horizontal extent which can be estimated as the radial position of the minimum in Fig.~\ref{fig:theory_sketch}\textit{B} (or Fig.~4A in the main text) as $r_{h} \approx 0.2R_0 \approx 116$ $\mu$m.

\subsection*{Some supporting results from the simulations} 

In this subsection we provide a number of results from the boundary integral simulations in support of the line of reasoning in the main document. The first result is the comparison of BI simulations of the impact of a HFE-7000 droplet in HFE-7000 vapor and in a hypothetical non-condensable gas that otherwise has the same properties as HFE-7000 vapor in equilibrium, only that it is in addition non-condensable. Whereas the former is modelled by a lubrication layer including vapor compressibility and phase change as discussed above, the latter is approximated using an incompressible lubrication layer formulation as was also used in \cite{Bouwhuis2012MaximalImpact}. We may justify this choice by estimating the threshold above which compressibility becomes important as $U_\text{thr,comp} \approx [P_{V,0}^3\mu_V/(\rho_L^4 R_0)]^{1/7} \approx 1.0$ m/s \cite{Mandre2009PrecursorsSurface}, such that the region of interest for this study lies in the incompressible regime.

In Fig.~\ref{fig:theory_sketch}\textit{B-F} we present the time evolution of the droplet profiles close to the substrate for five values of the impact speed $U_0 = 0.2$, $0.4$, $0.6$, $0.8$, and $1.0$ m/s, where in each panel the top plot corresponds to the HFE-7000 vapor case at $T = 22$ $^\circ$C (i.e., the first four plots are identical to those in Fig.~4A of the main document), and the lower plot to the corresponding hypothetical non-condensable gas. 

For the lowest impact velocity ($U_0 = 0.2$ m/s in Fig.~\ref{fig:theory_sketch}\textit{B}), the profiles are quite similar in both cases, although the dimple in the non-condensable case (bottom) is more stable than that found in the vapor case (top), where the center slowly moves down. Differences become more pronounced for $U_0 = 0.4$ m/s (Fig.~\ref{fig:theory_sketch}\textit{C}), where for the non-condensable gas (bottom) the dimple is slightly smaller in size (as expected from the results presented in \cite{Bouwhuis2012MaximalImpact}) but otherwise similar to that found for $U_0 = 0.2$ m/s. Note that the dimple is slowly moving up towards the end of the impact process, which can be understood from the color coding which serves to indicate the direction of increasing time from yellow to dark blue. For the vapor, however, the dimple, and the subsequently entrapped vapor bubble, is much smaller than for $U_0 = 0.2$ m/s, which also can be noted from the substantial downward motion of its center throughout the process.

For the non-condensable gas case (bottom), the slow decrease of the dimple size continues for the next velocity ($U_0 = 0.6$ m/s in Fig.~\ref{fig:theory_sketch}\textit{D}), but for the vapor (top) one observes a radical change: close to impact the bottom of the profile is almost flat with a minimum in the center such that no dimple is present and virtually no vapor is entrapped. And even if a small ring of vapor might be entrapped, it is likely quickly condensed due to the pressure rise immediately after impact.

The qualitative differences between the impact in vapor and non-condensable gas become even more clear cut for the last two velocities, $U_0 = 0.8$ and $1.0$ m/s in Figs.~\ref{fig:theory_sketch}\textit{E} and~\ref{fig:theory_sketch}\textit{F}: For the vapor case (top) there is hardly any sign of the presence of an intermediate phase, where the droplet hits the substrate almost without braking, whereas for the non-condensable gas case the expected gradual decrease of the dimple size continues. It is good to realize that the successive curves are equidistant in (dimensionless) time, at intervals $\tau = 1.33 \cdot10^{-3} R_0/U_0$, taking into account the impact velocity such that for a freely moving droplet successive curves would be equidistant in space as well, regardless of the value of $U_0$. This allows for assessing the relative deceleration in each case just by judging the distance between the profiles. Clearly, the large distance between the curves, e.g., found for $U_0 = 1.0$ m/s in the vapor case (Fig.~\ref{fig:theory_sketch}\textit{F}, top) indicates virtually undecelerated motion, where the liquid will hit the surface suddenly and hard when it arrives, whereas the bottom of the profile in the non-condensable gas case shows that in the center the liquid has come to a full stop, so to say in mid air, without any contact with the substrate, indicating a slow, cushioned landing.  

In the final panel, Fig.~\ref{fig:theory_sketch}\textit{G}, we compare the time evolution of the center height $h_c$, normalized with the starting height $h_0 = 0.02R_0$ of the simulation, for impact velocities $U_0$ varying between $0.2$ m/s and $1.0$ m/s in steps of $0.1$ m/s indicated by color changing from blue to yellow. In both cases, the straight dashed black line represents the time evolution of the position of the bottom of a moving sphere in the absence of vapor or gas. For the vapor (top plot, identical to Fig. 4\textit{B} in the main document), the dimple height decreases in time, a process that rapidly accelerates for larger $U_0$ and designates a non-decelerated, hard impact. In contrast, for the non-condensable gas the center height $h_c$ stabilizes in all cases, indicating the entrapment of a gas bubble below the droplet and a soft, cushioned landing.

To avoid possible confusion, it should be stressed that droplets are initiated at the starting height $h_0 = 0.02 R_0$ with a fixed velocity $U_0$, where $h_0$ has been chosen large enough to not influence the dynamics in the lubrication layer, but small enough to avoid  unnecessary computation time. The starting height is so small that gravity is irrelevant on that scale.\\ 

The final figure we have added, serves to elucidate the hyperbolic fit used to describe the data in Fig.~4\textit{C} in the main document. To this end, the latter is reproduced as Fig.~\ref{fig:num_fit}\textit{A}, plotting the center touchdown time $\Delta t$ (blue pentagons), estimated by extrapolating the time evolution of the center height $h_c$, as discussed in the main text, as a function of the impact velocity $U_0$. In Fig.~\ref{fig:num_fit}\textit{B} we plot the same data, but now with the reciprocal of the center touchdown time, $1/\Delta t$, on the vertical axis, from which it becomes clear that the data are reasonably well described by a linear fit
\begin{equation}\label{eq:linfit}
   \frac{1}{\Delta t} =  C (U_0 - U_{0,\text{thr}}) \quad\text{for}\,\,U_0 > U_{0,\text{thr}}\,\,,
\end{equation} 
with fitting parameters $U_{0,\text{thr}}$ and $C$, from which we find that $U_{0,\text{thr}}=0.50 \pm 0.02$ m/s. The continuous red curve in Fig.~\ref{fig:num_fit}\textit{A} is now of course just the reciprocal of \eqref{eq:linfit}, i.e.,
\begin{equation}\label{eq:hyperbola}
   \Delta t =  \frac{1}{C (U_0 - U_{0,\text{thr}})} \quad\text{for}\,\,U_0 > U_{0,\text{thr}}\,\,,
\end{equation} 
and the dashed red vertical line is its asymptote at $U_0 = U_{0,\text{thr}}$.

There is however also a new element in Fig.~\ref{fig:num_fit}\textit{A}, which is the estimated entrapped gas volume $V_\text{e,ng}$ for the case of the non-condensable gas simulations, denoted by the green triangles. Clearly, the entrapped non-condensable gas volume always remains appreciable, although it decreases monotonically with impact speed $U_0$ (consistent with \cite{Bouwhuis2012MaximalImpact}), whereas the entrapped vapor volume $V_\text{e}$ rapidly goes to zero at the threshold value $U_{0,\text{thr}}$. In the inset we show in a doubly logarithmic plot that the behavior of the entrapped volume $V_{e,nc}$ as a function of impact speed $U_0$ is consistent with the scaling law $V_{e,nc}/R_0^3 \sim \mathit{St}^{-4/3} \sim U_0^{-4/3}$ from \cite{Bouwhuis2012MaximalImpact}.

\section{Adding substrate superheat to the simplified model}

As discussed in the main text, the left hand side of the mass-energy balance equation Eq. (4) can be extended with a parasitic linear heat conduction term through the vapor layer, $Sk_V (T_s - T_V)/h$, with $T_s$ and $T_V$ the temperatures of the substrate and vapor, respectively, such that we obtain 
\begin{equation}\label{eq:massheatbalance1}
    -L \frac{dm_V}{dt} + Q_0 S k_V \frac{T_s - T_V}{h} =  K_0 S \, k_L \frac{\Delta T_V}{\delta_{th}}\,\,,
\end{equation} 
where $K_0$ and $Q_0$ are numerical constants of order one that serve to absorb the various approximations made in the modelling, $k_L$ and $k_V$ the thermal conductivity of the liquid and vapor phase, respectively, $L$ the latent heat of vaporization, $S$ the surface area of the liquid-vapor interface, $dm_V/dt$ the rate of change of vapor mass due to phase change, $\Delta T_V = T_V - T_0$ the difference between the vapor temperature and the ambient temperature $T_0$, and $\delta_{th}$ the thermal boundary layer thickness in the liquid. Decomposing $T_s - T_V = (T_s - T_0) - (T_V - T_0)$ we obtain
\begin{equation}\label{eq:massheatbalance2}
    -L \frac{dm_V}{dt} + Q_0 S k_V \frac{T_s - T_0}{h} =  S \, 
   \left[K_0\frac{k_L}{\delta_{th}} + Q_0\frac{k_V}{h}\right]\Delta T_V \,\,.
\end{equation} 
Since $k_L$ is almost an order of magnitude larger than $k_V$ (see Table~\ref{table:NovecProp}) and since $h$ is (at least up to the dimple formation) typically larger than $\delta_{th}$, we can neglect the second term on the right hand side with respect to the first, such that
\begin{equation}\label{eq:massheatbalance3}
    -L \frac{dm_V}{dt} + Q_0 S k_V \frac{\Delta T}{h} = K_0 S \, k_L \frac{\Delta T_V}{\sqrt{\pi\alpha_L \Delta t}}\,\,,
\end{equation} 
where we have used that $\delta_{th} \approx \sqrt{\pi\alpha_L \Delta t}$ with $\alpha_L$ the thermal diffusivity, and $\Delta t \approx \epsilon R_0/U_0$ with $\epsilon = 0.02$ the time span during which the pressure rise in the lubrication layer is appreciable. Here, $R_0$ and $U_0$ are the radius and velocity of the impacting droplet. If the superheat $\Delta T = T_s - T_0$ of the substrate equals zero, we retrieve Eq. (4) from the main text. We now follow the same steps as in Eqs.~(4) to~(8) in the main text, introducing the speed of the condensation front in the vapor phase $U_{cond} = -(dm_V/dt)/(\rho_V S)$, with $\rho_V$ the vapor density, which then leads to
\begin{equation}\label{eq:massheatbalance4}
    \frac{U_{cond}}{U_0} + Q_0  \frac{k_V}{\rho_V L U_0} \frac{\Delta T}{h} =  K_0  \, \frac{\alpha_L\rho_Lc_{p,L}}{\rho_V L U_0} \frac{\Delta T_V}{\sqrt{\pi\alpha_L \Delta t}}\,\,,
\end{equation} 
where we have used that $k_L = \alpha_L\rho_Lc_{p,L}$, with $\rho_L$ and $c_{p,L}$ the density and specific heat of the vapor phase. If, again, we use the linearized Clausius-Clapeyron equation (Eq. (5) in the main text) to connect $\Delta T_V$ to $\Delta P_V$, we obtain
\begin{equation}\label{eq:massheatbalance5}
    \frac{U_{cond}}{U_0} + Q_0  \frac{k_V}{\rho_V L U_0} \frac{\Delta T}{h} =  \frac{K_0\beta^2}{\sqrt{\pi\epsilon}} \frac{\rho_L}{\rho_V}\frac{c_{p,L}}{R_s}\sqrt{\frac{\alpha_L}{R_0 U_0}} \frac{\Delta P_V}{P_{V,0}}\,\,,
\end{equation} 
which is the equivalent of Eq. (6) from the main text to which it reduces for $\Delta T = 0$. Remember that $\beta = R_sT_0/L$, with $R_s$ the specific gas constant of the vapor.  Now, finally, we may use Eq. (1) from the main text to estimate $\Delta P_V \sim \mu_V R_0U_0/h^2$ which then leads to 
\begin{equation}\label{eq:massheatbalance6}
    \frac{U_{cond}}{U_0} + Q_0  \frac{k_V}{\rho_V L U_0} \frac{\Delta T}{R_0}\frac{R_0}{h} =  \frac{K_0\beta^2}{\sqrt{\pi\epsilon}} \frac{\rho_L}{\rho_V}\frac{c_{p,L}}{R_s}\sqrt{\frac{\alpha_L}{R_0 U_0}} \frac{\mu_V}{\rho_L R_0 U_0}\frac{\rho_LU_0^2}{P_{V,0}}\frac{R_0^2}{h^2}\,\,,
\end{equation}
where we have conveniently introduced dimensionless groups in which one recognizes the Euler, Stokes and P{\'e}clet numbers $\mathit{Eu} = \rho_LU_0^2/P_{V,0}$, $\mathit{St} = \rho_L R_0 U_0/\mu_V$ and $\mathit{Pe} = R_0U_0/\alpha_L$. Moreover we can now use the dimple formation condition Eq. (2) from the main text, $h_d/R_0 \sim \mathit{St}^{-2/3}$ to eliminate the remaining unknown layer thickness $h$, using the same argumentation as provided in the main text, namely that one needs to evaluate the ratio $U_{cond}/U_0$ at the moment that the dimple would be formed in a non-condensable gas with otherwise the same properties as the vapor, i.e., at $h=h_d$.     

For zero superheat, $\Delta T = 0$, this of course leads to the same Eq. (8) as we found in the main text, which we repeat below for multiple reasons
\begin{eqnarray}
\left.\frac{U_{cond}}{U_0}\right|_{\Delta T = 0} &=& \frac{K_0\beta^2}{\sqrt{\pi\epsilon}}\frac{\rho_L}{\rho_{V,0}}\frac{c_{p,L}}{R_s}\frac{\mathit{Eu}\,\mathit{St}^{1/3}}{\mathit{Pe}^{1/2}} \nonumber\\
&=& \frac{K_0\beta^2}{\sqrt{\pi\epsilon}}\frac{\rho_L}{\rho_{V,0}}\frac{c_{p,L}}{R_s} \frac{\alpha_L^{1/2}\rho_L^{4/3}U_0^{11/6}}{\mu_V^{1/3}P_{V,0}R_0^{1/6}} = \left[\frac{U_0}{U_0^*}\right]^{11/6}\,\,.
\label{eq:finalDT0}
\end{eqnarray} 
The first is to show how the constant $K_0$, which absorbs all of the uncertainties in the approximations made along the way, enters the equation, the second is to manifestly show that the right hand side is proportional to $U_0^{11/6}$, and the third is to introduce a velocity scale $U_0^*$, which from the previous step can be straightforwardly shown to be equal to
\begin{equation}
U_0^* \equiv  \left[\frac{\sqrt{\pi\epsilon}}{K_0\beta^2}\frac{\rho_{V,0}}{\rho_L}\frac{R_s}{c_{p,L}} \frac{\mu_V^{1/3}P_{V,0}R_0^{1/6}}{\alpha_L^{1/2}\rho_L^{4/3}}\right]^{6/11}
\label{eq:U0star}
\end{equation}
Clearly, $U_0^*$ represents the threshold impact velocity for which the condition $U_{cond}/U_0|_{\Delta T = 0} = 1$ is satisfied. (Note that in Eq.~\eqref{eq:finalDT0} we have replaced $\rho_V \approx \rho_{V,0}$ in leading order.)

For finite $\Delta T$ this definition somewhat simplifies our result \eqref{eq:massheatbalance6} into
\begin{equation}\label{eq:massheatbalance7}
    \frac{U_{cond}}{U_0} + Q_0  \frac{k_V T_0 \rho_L^{2/3} }{\rho_{V,0} L \mu_V^{2/3} R_0^{1/3}(U_0^*)^{1/3}} \frac{\Delta T}{T_0}\left[\frac{U_0}{U_0^*}\right]^{-1/3} =  \left[\frac{U_0}{U_0^*}\right]^{11/6}\,\,,
\end{equation}
where, again we used the dimple condition to eliminate $h$ from the second term on the left hand side to explicitly show its dependence on the impact speed $U_0$. Applying the condition $U_{cond}/U_0 = 1$ to find the boundary between the region in which vapor bubble entrapment occurs and the region where there is no entrapment, we obtain from the above equation
\begin{eqnarray}\label{eq:boundary}
    \frac{\Delta T}{T_0} &=&  \frac{1}{Q_0}  \frac{\rho_{V,0} L \mu_V^{2/3} R_0^{1/3}(U_0^*)^{1/3}}{k_V T_0 \rho_L^{2/3} }\left(\left[\frac{U_0}{U_0^*}\right]^{13/6} - \left[\frac{U_0}{U_0^*}\right]^{1/3}\right) \nonumber\\
    &=& X\left(\left[\frac{U_0}{U_0^*}\right]^{13/6} - \left[\frac{U_0}{U_0^*}\right]^{1/3}\right)\,\,,
\end{eqnarray}
where we defined
\begin{equation}
X \equiv \frac{1}{Q_0}\,\frac{\rho_{V,0} L \mu_V^{2/3} R_0^{1/3}(U_0^*)^{1/3}}{k_V T_0 \rho_L^{2/3} } = \frac{1}{Q_0}\frac{1}{\beta}\frac{R_s}{c_{p,V}}\left[\frac{\mu_V}{\rho_LR_0U_0^*}\right]^{2/3}\frac{R_0U_0^*}{\alpha_V}\,\,.
\label{eq:X}
\end{equation}
To obtain the last expression we used $k_V = \alpha_V \rho_{V,0} c_{p,V}$, with which one recognizes the Stokes number $\mathit{St}^* = \rho_LR_0U_0^*/\mu_V$ and vapor P{\'e}clet number $\mathit{Pe}_V^* = R_0U_0^*/\alpha_V$, both based on the threshold velocity $U_0^*$. 

It is \eqref{eq:boundary} that describes the boundary between the regions where entrapment occurs and where there is no entrapment. Inserting the droplet radius $R_0 = 0.6$ mm and the physical properties of HFE-7000, the equation has two remaining fitting parameters, $K_0$ and $Q_0$, from which the first is fixed to 
$K_0 = 1.6$ to obtain an appropriate value $U_0^* = 0.52$ m/s for the threshold velocity at zero superheat, and the second one is determined as 
$Q_0 = 1.35$ leading to a fair description of the boundary visible in the experimental data of Fig. 5 of the main text, as evidenced from the blue curve in the phase diagram.  

\begin{figure}
\centering
\captionsetup{labelformat=empty}
\caption*{\emph{[Continued from previous page:]}\\
Liquid HFE-7000 is stored in a reservoir Res1 to ensure saturation at temperature $T_{0}$. Inside the vessel, drops are generated from a second HFE-7000 reservoir Res2 connected to the needle N via valve V4. The drops with equilibrium radius $R_0$ and velocity $U_0$ impact a sapphire prism P with temperature $T_s$, controlled by a resistive heater and a PT-100 sensor (not pictured). Additionally, a sensor reads the absolute pressure and PT-100 sensors monitor the temperatures at the chamber walls and reservoir Res 1. We record side and bottom views of the impact. Side views are backlit with a white collimated light source CLS1 and imaged with a long distance microscope LDM attached to a high speed camera CAM1.  For the bottom views, frustrated total internal reflection imaging (TIR) is used to distinguish wet and dry zones of the surface. A blue ($\sim 455$ nm) collimated light beam from the LED source CLS2 is totally reflected by the prism (at an angle greater than the critical angle for sapphire), unless a droplet starts to wet the surface and allows light to transmit instead, resulting in a shadowgraphy of the wet and dry sections of the surface. An image of the surface is generated with an infinity corrected microscope objective ICO inside the setup which passes the light beam via a  mirror M into the high speed camera CAM2 with a tube lens TL attached.\\ 
(\textit{B}) TIR image sequence (after background subtraction) of a HFE-7000 droplet with impact velocity $U_0 = 0.38$  m/s and radius $R_0=580 \, \mu$m in a vapor atmosphere at $T_{0} = 22.7 \, ^\circ$C, where the droplet impacts a sapphire substrate heated to a temperature difference  $\Delta T = 1.3 \, ^\circ$C. Under these conditions a vapor bubble is entrapped below the droplet. A dark ring can be seen 
in the first frame, indicating that the droplet starts wetting within the first 5 $\mu$s of impact but it also entraps vapor (the dry zone within the ring). Over the next several frames the outer contact line expands, while the inner retracts. In the last frames a small dot can be seen in the center of the still entrapped vapor bubble. Soon after the last image the entrapped vapor is no longer visible in the TIR images.}
\end{figure}

\begin{sidewaysfigure}
    \includegraphics[width=0.8\linewidth]{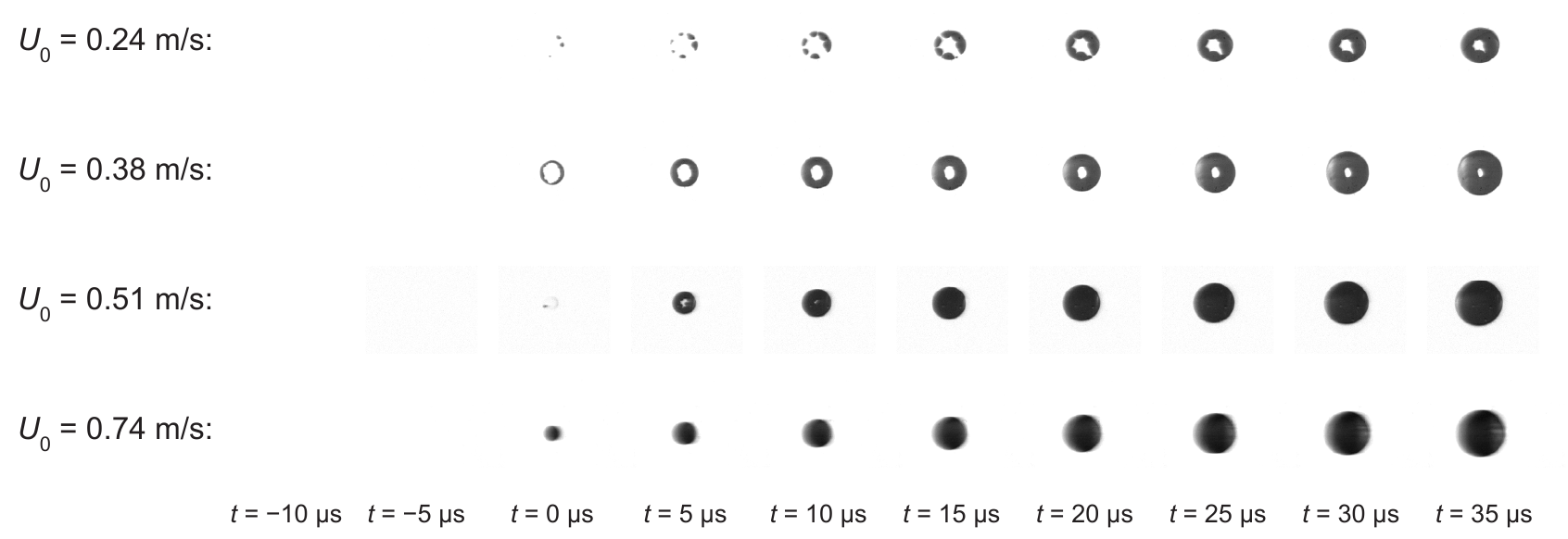}
\caption{\label{fig:extraimages} TIR image sequence (after background subtraction) of a series of HFE-7000 droplets with radius $R_0 \approx 0.6$ mm in a vapor atmosphere at $T_{0} \approx 22$ $^\circ$C; the droplets impact a sapphire substrate heated to a temperature difference  $\Delta T \lesssim 2$ $^\circ$C with impact velocities $U_0 = 0.24$, $0.38$, $0.51$, $0.74$ m/s, Movies S1 to S5 in this SI. At low impact velocity a vapor bubble is entrapped, but at high impact velocities this entrapment is suppressed. Note that this is an extended version of Fig. 2\textit{A} in the main text, which includes additional reaspected snapshots.}
\end{sidewaysfigure}

\begin{figure}
    \centering
    \includegraphics[width=0.95\linewidth]{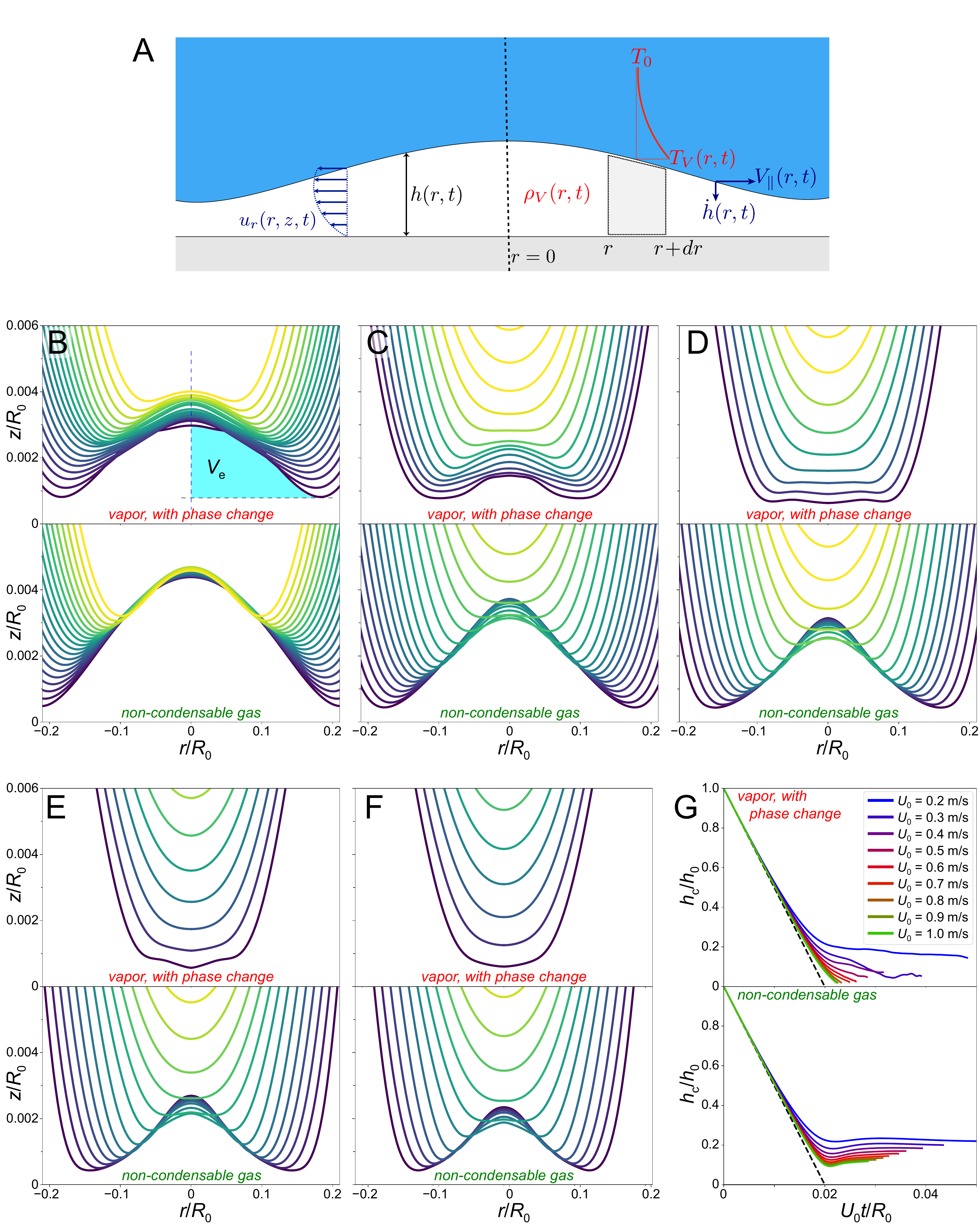}
    \caption{\label{fig:theory_sketch} (\textit{A}) Sketch of an impacting droplet. Top panels in \textit{B}-\textit{G} correspond to the impact of a HFE-7000 droplet in boiling liquid conditions (i.e., in thermal equilibrium with its vapor), bottom panels to a HFE-7000 droplet in a non-condensable gas, with otherwise the same properties as HFE-7000 vapor. Below the droplet there is a radial flow $u_r(r,z,t)$ of vapor (or gas in the lower plots) due to its proximity to the substrate, which pressurizes the vapor (or gas) layer. In the top plots this triggers condensation, with the latent heat flowing into the liquid through the thermal boundary layer that is provoked by the pressure rise. In the bottom plots there is no phase change and non-condensable gas is entrapped below the droplet.\\ 
\emph{[Caption continued on next page:]}}
\end{figure}

\begin{figure}
\centering
\captionsetup{labelformat=empty}
\caption*{\emph{[Continued from previous page:]}\\
(\textit{B}-\textit{F}) Comparison of droplet profiles from BI simulations of the impact of a HFE-7000 droplet in HFE-7000 vapor, modelled by a lubrication layer including vapor compressibility and phase change (top plots), and BI simulations with the vapor replaced by a non-condensable gas with exactly the same properties as the HFE-7000 vapor, using an incompressible lubrication layer (bottom plots). We present data for (\textit{B}) $U_0 = 0.2$ m/s, (\textit{C}) $U_0 = 0.4$ m/s, (\textit{D}) $U_0 = 0.6$ m/s, (\textit{E}) $U_0 = 0.8$ m/s, and (\textit{F}) $U_0 = 1.0$ m/s. Note that the upper plots of (\textit{B}-\textit{E}) are identical to those presented in the main text in Fig.~4\textit{A}. It is clear that, whereas for the non-condensable gas case a gas pocket is entrapped in all cases, with the pocket becoming smaller with increasing velocity $U_0$, consistent with \cite{Bouwhuis2012MaximalImpact}, for the vapor case a vapor pocket is entrapped for $U_0 = 0.2$ and $0.4$ m/s only, and for the other cases the droplet hits the substrate with a strongly diminished cushioning effect. All plots are on the same scale, and the time interval $\tau$ between successive curves have been normalized with velocity, such that  $\tau = 1.33 \cdot10^{-3} R_0/U_0$. 
(\textit{G}) Comparison of the time evolution of the center height $h_c$, normalized with the initial height $h_0$, for various impact velocities $U_0$ between $0.2$ m/s and $1.0$ m/s (blue to yellow), for the vapor case (top) and the non-condensable gas case (bottom). In both cases, the straight dashed black line represents the time evolution of the position of the bottom of a moving sphere in the absence of vapor or gas. Whereas for the non-condensable gas the center height $h_c$ stabilizes in all cases, for the vapor the center height decreases in time, a process that rapidly accelerates for larger $U_0$.
}
\end{figure}
\clearpage
\newpage
\begin{figure}
    \centering
     \includegraphics[width=1\linewidth]{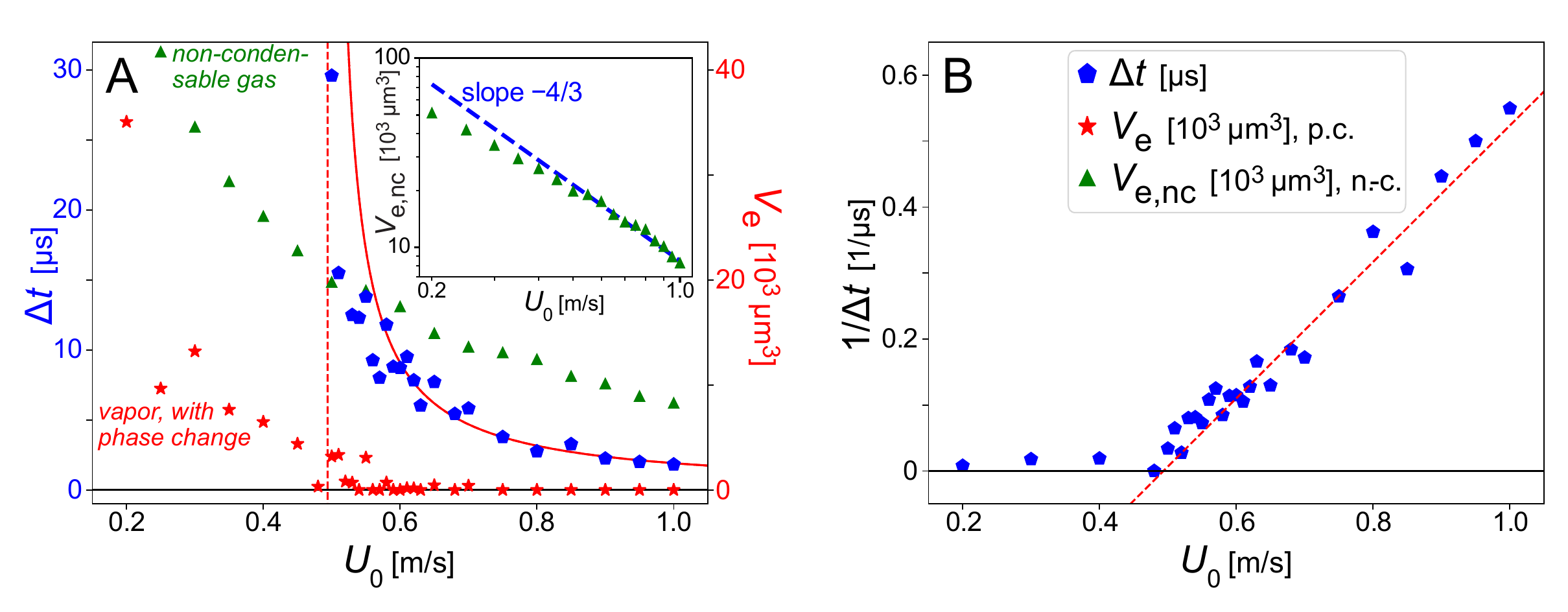}
    \caption{(\textit{A}): The center touchdown time $\Delta t$ (blue pentagons), plotted as a function of the impact velocity $U_0$, where the continuous red curve is a hyperbolic fit to the data and the dashed red vertical line is its asymptote. The red stars represent the entrapped vapor volume $V_\text{e}$, estimated from the volume enclosed inside and above the minimum of the last profile before touchdown (see Fig.~\ref{fig:theory_sketch}\textit{A} (or Fig.~4\textit{A} in the main document), where it is the volume of revolution created by revolving the lightblue area indicated in the $U_0 = 0.2$ m/s plot around the vertical axis). If the minimum is in the center ($r = 0$) the entrapped volume is zero. The green triangles denote the entrapped non-condensable gas volume $V_\text{e,nc}$ obtained for the non-condensable simulations also presented in Fig.~\ref{fig:theory_sketch} and are the only new element added to the plot provided in Fig.~4\textit{C} in the main text. The inset compares the entrapped non-condensable gas volume $V_\text{e,nc}$ to the expected power-law behavior $V_{e,nc} \sim \mathit{St}^{-4/3} \sim U_0^{-4/3}$ in a doubly logarithmic plot (see Ref.~\cite{Bouwhuis2012MaximalImpact}). Deviations for smaller $U_0$ are attributed to the proximity of the maximum discussed in that work.
(\textit{B}): Reciprocal of the center touchdown time, $1/\Delta t$, plotted against the impact velocity $U_0$. The red dashed straight line is a linear fit to the non-zero data ($< 0.03$ $\mu$s$^{-1}$, which corresponds to the hyperbola in \textit{A}. Note that the legend in \textit{B} applies to the data in both figures.}
    \label{fig:num_fit}
\end{figure}
\clearpage

\newpage

\bibliographystyle{prsty_withtitle}
\bibliography{references_mendeley_ext}